\documentclass[a4paper,twocolumn,11pt]{quantumarticle}
\pdfoutput=1
\usepackage[utf8]{inputenc}
\usepackage[english]{babel}
\usepackage[T1]{fontenc}
\usepackage{amsmath, amsthm, amssymb, amsfonts}
\usepackage{physics, bm}
\newcommand\sbullet[1][.5]{\mathbin{\vcenter{\hbox{\scalebox{#1}{$\bullet$}}}}}
\usepackage[numbers,sort&compress]{natbib}
\usepackage{tikz}
\usepackage{lipsum}
\usepackage{subcaption}
\usepackage{float}
\usepackage{textgreek}
\usepackage{url} 
\usepackage[pdftex, unicode]{hyperref}
\usepackage{bookmark}
\makeatletter
\providecommand{\@afterenddocumenthook}{}
\makeatother

\usepackage{silence}
\makeatletter
\def\ignore@rerun@check{}
\makeatother

\begin{document}

\title{Quantum Error Mitigation at the pre-processing stage}

\author{Juan F. Martín}
\orcid{0009-0005-7025-2347}
\author{Giuseppe Cocco}
\orcid{0000-0003-1720-8986}
\author{Javier Fonollosa}
\orcid{0000-0002-0136-2586}
\affiliation{Department of Teoria del Senyal i Comunicacions, Universitat Politècnica de Catalunya, ES-08034 Barcelona, Spain}
\maketitle

\begin{abstract}
The realization of fault-tolerant quantum computers remains a challenging endeavor, forcing state-of-the-art quantum hardware to rely heavily on noise mitigation techniques. Standard quantum error mitigation is typically based on post-processing strategies. In contrast, the present work explores a pre-processing approach, in which the effects of noise are mitigated before performing a measurement on the output state. The main idea is to find an observable $Y$ such that its expectation value on a noisy quantum state $\mathcal{E(\rho)}$ matches the expectation value of a target observable $X$ on the noiseless quantum state $\rho$. Our method requires the execution of a noisy quantum circuit, followed by the measurement of the surrogate observable $Y$. The main enablers of our method in practical scenarios are Tensor Networks. The proposed method improves over Tensor Error Mitigation (TEM) in terms of average error, circuit depth, and complexity, attaining a measurement overhead that approaches the theoretical lower bound. The improvement in terms of classical computation complexity is in the order of \( \sim 10^6 \) times when compared to the post-processing computational cost of TEM in practical scenarios. Such gain comes from eliminating the need to perform the set of informationally complete positive operator-valued measurements (IC-POVM) required by TEM, as well as any other tomographic strategy.
\end{abstract}

\section{Introduction}

One of the main research objectives in quantum computing is the realization of fault-tolerant quantum computers. One of the most promising approaches is the use of error-correcting codes that suppress noise below a critical threshold, enabling reliable quantum operations \cite{preskillbook}. While Quantum Error Correction (QEC) \cite{Roffe2019, Chatterjee2023} has been a major focus of the community over the past decade \cite{Campbell2024, Acharya2024}, its practical implementation remains challenging due to the considerable overhead required in terms of qubits. Current state-of-the-art hardware falls far short of the hundreds to millions of logical qubits needed for scientific \cite{Kivlichan2020} and industrial \cite{Lee2021} applications, leaving fault-tolerant quantum computing a distant prospect. Current noisy intermediate-scale quantum (NISQ) regime \cite{Preskill2018} is characterized by large error rates and a limited qubit count, which hinder practical quantum computing. Although recent experimental advancements in early-stage fault-tolerant capabilities offer promising glimpses of progress toward bridging this gap \cite{Bedalov2024, Singh2024, Acharya2023, Takeda2022, Postler2022, Krinner2022, Egan2021, Abobeih2022, Ryan2022, Malone2022, Abughanem2024}, Quantum Error Mitigation (QEM) \cite{cai2023, Kandala2019} remains the best countermeasure so far.

In this context, QEM techniques have emerged as a practical alternative, serving as a bridge between NISQ devices and fault-tolerant quantum computing. QEM aims to suppress noise in medium-depth quantum circuits through repeated executions and statistical post-processing of measurement data. While QEC remains the most promising long-term solution, QEM provides a viable pathway to harness the potential of current quantum hardware, offering hope for achieving quantum advantage before the full realization of fault-tolerant systems \cite{Suzuki2022}.

Early QEM strategies were designed to operate without specific knowledge of the underlying noise processes. Examples are Zero Noise Extrapolation (ZNE) \cite{Temme2017, Li2017}, Virtual Distillation \cite{Huggins2021}, Symmetry Expansion \cite{Cai2021} and Subspace Expansion \cite{Yoshioka2022}. Recent advancements in quantum technology have enabled precise characterization of noise in quantum circuits (e.g. the noise associated with the application of each unitary gate). Such knowledge allowed to develop more efficient error mitigation approaches. Notable examples are QEM with Artificial Neural Networks \cite{Kim2020}, Zero Noise Extrapolation with probabilistic error amplification (ZNE-PEA) \cite{Kim2023}, Probabilistic Error Cancellation (PEC) \cite{ewout2023}, Restricted Evolution (EMBRE) \cite{Saxena2024} and Tensor Error Mitigation (TEM) \cite{Filippov2023}.

\subsection{Contributions}
Given a noiseless quantum state $\rho$ and target observable $X$, the mitigation of the resulting expected value $\Tr{\rho X}$ requires, for most existing mitigation techniques, multiple shots across a big set of quantum circuits. An alternative is to construct a \textit{surrogate} observable \(Y\) such that its expectation value on the noisy quantum state \(\mathcal{E}(\rho)\) coincides with the expectation value of the target observable \(X\) on the ideal state \(\rho\). This problem was first explored by Watanabe \textit{et al.} \cite{Watanabe2010}. The main obstacle to this approach, as mentioned in Ref. \cite{Watanabe2010}, is that obtaining $Y$ requires keeping track of an exponential number of terms. In this paper, we revisit this initial idea, studying the evolution of the original observable $X$ through a specific quantum channel, which is related to the Heisenberg evolution of the \textit{surrogate} observable $Y$, and characterized by a sparse Pauli-Lindblad noise description. To achieve this, we make use of state-of-the-art unbiased techniques provided by the Tensor Network (TN) representation. Inspired by the recent addition of TEM \cite{Filippov2023} to the set of QEM methods, we demonstrate that it is indeed possible to find such an observable $Y$ in practical setups. We show that such an observable saturates the Quantum Cramér-Rao Bound (QCRB) \cite{Braunstein1994}, yielding an optimal estimator with minimal variance.

We also show that, in the case where the original observable is a single Pauli string \(X = P_i\), the optimal estimator \(\hat{Y}\) can be well approximated by a rescaled version of the same Pauli operator. Our simulations indicate that such a simple approximation, which we denote as Dominant Component Approximation (DCA), improves expectation value estimation over TEM in terms of error, variance, and time complexity, saving up to $\sim 10^6$ TN contractions. For such a comparison, we consider the evolution of a quantum state through the Trotterization of the Ising model. Besides, our approach eliminates the need for additional quantum operations, such as shadow tomography, required by TEM, so no prior information about the original (noisy or noiseless) quantum state is required.

\begin{figure*}
    \centering
    \includegraphics[width=1.0\linewidth]{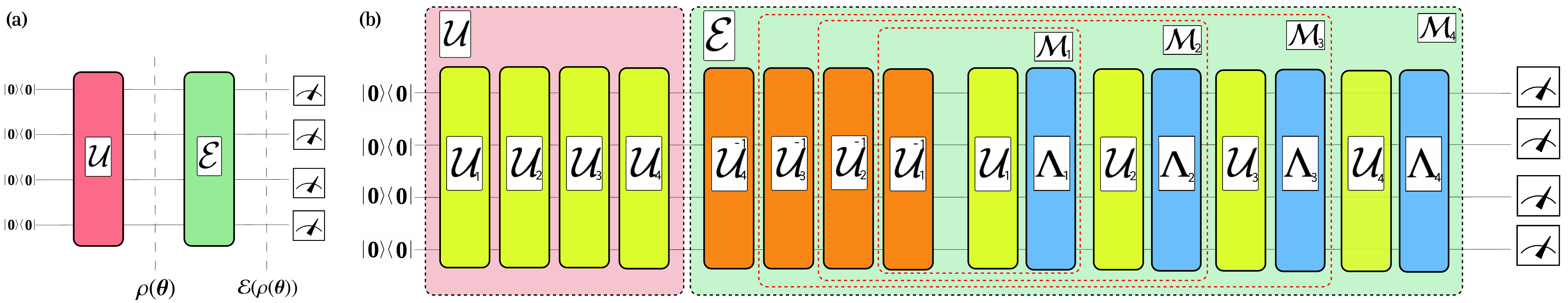}
    \caption{\footnotesize Representation of our noisy circuit model. (a) One-layer circuit, where $\mathcal{U}$ is a (one-layer) unitary transformation; $\rho(\theta) = \mathcal{U}(\ketbra{0}^{\otimes n})$ is the ideal output of the noiseless quantum circuit; $\mathcal{E}(\sbullet)$ is the (one-layer) channel used to model the noise that affects the ideal quantum state; $\mathcal{E}(\rho)$ is the noisy quantum state at the output of the noisy quantum circuit. (b) Multi-layer example of 4 unitary layers. $\mathcal{U}_l$ indicates the $l$-th unitary layer, such that $\mathcal{U} = \bigcirc_{l}\mathcal{U}_l$, and the noiseless quantum state is $\rho(\theta) = \mathcal{U}(\ketbra{0}^{\otimes n}$); $\mathcal{E}(\sbullet)$ is the (multi-layer) noise channel composed of noiseless unitary layers $\mathcal{U}_l$, their inverses $\mathcal{U}_l^{-1}$ and the one-layer noise map $\Lambda_l$. The dotted red lines indicate the contraction step $l$ for the middle-out contraction matrix $\mathcal{M}_l$.}
    \label{fig: circuit_paint}
\end{figure*}

\section{Overview}
In this section, we propose the theoretical framework underlying our work, along with a description of the techniques employed. Section \ref{sec: problem description} reviews the original problem from an analytical perspective, Section \ref{sec: noisy computation} provides a detailed characterization of the noise model, Section \ref{sec: TN approach} discussed the Tensor Network (TN) formalism, while Section \ref{sec: surrogate} presents the derivation of the \textit{surrogate} observable $\hat{Y}$ based on TN.

\subsection{Problem description}
\label{sec: problem description}
Let $\mathcal{D}(\mathcal{H})$ denote the space of density operators $\rho$ acting on a Hilbert space $\mathcal{H}$; let $\mathcal{L}(\mathcal{H})$ denote the space of square linear operators acting on $\mathcal{H}$; and let $\mathcal{L}(\mathcal{H}_A, \mathcal{H}_B)$ denote the space of linear operators taking a Hilbert space $\mathcal{H}_A$ to a Hilbert space $\mathcal{H}_B$ of possibly different dimensions. A quantum state $\rho(\bm{\theta}) \in \mathcal{D}(\mathcal{H})$ of an $n$-qubit system and a traceless observable $X \in \mathcal{L}(\mathcal{H})$ can be decomposed into a combination of Pauli matrices as
\begin{align}
    \rho(\bm{\theta}) &= \frac{1}{2^n} \left(I_{2^n} + \sum_{i=1}^{4^n-1} \theta_i P_i \right) \nonumber \\
    &= \frac{1}{2^n} \left(I_{2^n} + \bm{\theta}^T \cdot \bm{P} \right),
        \label{eq: rho definition}
\end{align}
\begin{equation}
        X = \sum_{i=1}^{4^n-1} x_i P_i = \bm{x}^T \cdot \bm{P},
        \label{eq: def X}
\end{equation}
where $I_{2^n}$ is the identity matrix $I^{\otimes n}$ of dimension $2^n \times 2^n$, $\bm{\theta} \in \mathbb{R}^{4^n-1}$ is the $n$-qubit Bloch vector defining the quantum state $\rho$, $\bm{x} \in \mathbb{R}^{4^n-1}$ is an $n$-qubit vector that characterizes the observable $X$ and $\bm{P} = \{P_i\}_{i=1}^{4^n-1}$ is a vector of arrays composed by $n$-qubit Pauli strings excluding the trivial identity operator, i.e. $P_i \in \{\sigma_0, \sigma_1, \sigma_2, \sigma_3\}^{\otimes n} \setminus \{\sigma_0\}^{\otimes n}$, with $\sigma_0, \sigma_1, \sigma_2, \sigma_3$ being the single-qubit Pauli matrices of dimension $2 \times 2$. 

\setcounter{footnote}{0}

When running a quantum circuit, we are often interested in finding the expectation value of an observable $X$ when applied to an ideal noiseless output state $\rho(\bm{\theta})$, that is:
\begin{equation}
    \left<X\right>_{\rho(\bm{\theta})} = \Tr{\rho(\bm{\theta}) X} = \bm{\theta}^T \bm{x}.
    \label{eq: exp X}
\end{equation}
Eq. (\ref{eq: exp X}) follows from the facts that $\Tr{P_i} = 0$ and $\Tr{P_i P_j} = 2^n \delta_{ij}$ for $i,j \in \{1, \cdots, 4^n-1\}$. Current state-of-the-art quantum devices are affected by relatively strong noise. Thus, the actual quantum state that the circuit outputs might significantly differ from $\rho(\bm{\theta})$. To approach this problem, we describe a noisy unitary operation as the concatenation $(\mathcal{E} \circ \mathcal{U}) (\sbullet) $, where $\mathcal{U}$ represents the noiseless unitary transformation $\mathcal{U}(\sbullet) = U \sbullet U^\dagger$ with $U \in \mathcal{L}(\mathcal{H}, \mathcal{H})$, while $\mathcal{E}$ is a Completely-Positive Trace-Preserving (CPTP) map that models the noise experienced in the NISQ quantum processor when $\mathcal{U}$ is executed. Therefore, one may describe the impact of any noise over a quantum system as the action of a quantum channel $\mathcal{E}(\sbullet): \mathcal{L}(\mathcal{H})\xrightarrow{} \mathcal{L}(\mathcal{H})$. Fig. \ref{fig: circuit_paint}a provides a visual representation of such a noise model. Using this, the noisy quantum state can be expressed as
\begin{align}
        \mathcal{E}(\rho(\bm{\theta})) &= \frac{1}{2^n} \left( \mathcal{E}(I_{2^n}) + \sum_{i=1}^{4^n-1} \theta_i \mathcal{E}(P_i) \right) \\
        & = \rho(A \bm{\theta} + \bm{c}),
        \label{eq: 1}
\end{align}
where $A_{ij} = \frac{1}{2^n}\Tr{P_i \mathcal{E}(P_j)}$ and $c_i = \frac{1}{2^n} \Tr{P_i \mathcal{E}(I_{2^n})}$ \cite{Watanabe2010}. This shows that any CPTP noisy channel can be characterized by identifying a matrix $A$ and a vector $\bm{c}$\footnote{Note that in the special case where the noise channel is unital, $\mathcal{E}(I_{2^n}) = I_{2^n}$, then $c = 0$.}, as in Eq. (\ref{eq: 1}).

Let us now consider another general observable $Y \in \mathcal{L}(\mathcal{H})$, not necessarily traceless, of the form
\begin{equation}
        Y = y_0 I_{2^n} + \sum_{i=1}^{4^n-1} y_i P_i = y_0 I_{2^n} + \bm{y}^T \cdot \bm{P},
        \label{eq: Y in paulis}
\end{equation}
where $y_0 \in \mathbb{R}$ and $\bm{y} \in \mathbb{R}^{4^n -1}$. Its expectation value applied to the noisy state $\mathcal{E}(\rho(\bm{\theta}))$ has the form
\begin{align}
        \left<Y\right>_{\mathcal{E}(\rho(\bm{\theta}))} &= \Tr{\mathcal{E}(\rho(\bm{\theta})) Y} \label{eq: trace of Y} \\
        & = (\bm{A\theta} + \bm{c})^T \cdot \bm{y} + y_0.
\end{align}

According to the Heisenberg formalism, we can evolve the observable $Y$ through the adjoint quantum channel $\mathcal{E^\dagger(\sbullet)}$, obtaining an alternative expression for Eq. (\ref{eq: trace of Y}):
\begin{equation}
    \Tr{\mathcal{E}(\rho(\bm{\theta})) Y} = \Tr{\rho(\bm{\theta}) \mathcal{E}^\dagger(Y)}.
    \label{eq: adjoint channel}
\end{equation}
Our main objective is to find an accurate approximation for $Y$ such that
\begin{equation}
    \mathcal{E}^\dagger(\hat{Y}) = X \Longrightarrow \left<\hat{Y}\right>_{\mathcal{E}(\rho(\bm{\theta}))} \approx \left<X\right>_{\rho(\bm{\theta})}.
    \label{eq: final objective}
\end{equation}
Measuring the surrogate observable $\hat{Y}$ over the noisy quantum circuit allows us to extract noise-mitigated statistics. This approach avoids using complex post-processing algorithms with noisy data and eliminates the need to extract information from the quantum state via tomographic methods, which are generally resource-intensive and challenging to implement in practice.

An analytical treatment of this problem has been previously presented in~\cite{Watanabe2010}, yielding the expression
\begin{equation}
    \bm{y} = A^{-T} \bm{x}, \quad y_0 = - \bm{c}^{-T} A^{-T} \bm{x},
\end{equation}
which also holds for non-traceless observables \(X\), in which case the scalar term \(y_0\) is appropriately redefined. However, this solution is not directly applicable in practical settings, as it requires constructing and inverting the matrix \(A\), which has dimension \((4^n-1) \times (4^n - 1)\), thus becoming computationally intractable as the system size grows.

In this work, we build upon the theoretical foundation of Ref. \cite{Watanabe2010} and develop a scalable and efficient approximation strategy to compute and implement $\hat{Y}$. Specifically, we consider the Trotterization of the Ising model, often used to test QEM methods, and we employ the Tensor Network (TN) formalism to capture the structure of \(A\) in a compressed form.

The obtained \textit{surrogate} observable \(\hat{Y}\) is optimal in the sense that it achieves the Quantum Cramér-Rao bound (QCRB) \cite{Braunstein1994}, which establishes a fundamental lower bound on the variance for any estimator of \(\langle X \rangle_{\rho(\bm{\theta})}\) from $\mathcal{E}(\rho(\bm{\theta}))$. A detailed analysis of the QCRB for such estimator \(\hat{Y}\) is provided in Appendix \ref{app: crb}.  

\subsection{Noisy implementation}
\label{sec: noisy computation}

Most quantum systems of practical interest can be modeled as multiple-layer quantum circuits. The schemes in Fig. \ref{fig: circuit_paint}a and Fig. \ref{fig: circuit_paint}b represent a single and a multiple layer setup, respectively. The multiple-layer error model $\mathcal{E}(\sbullet)$ applied to the ideal quantum state $\rho(\bm{\theta})$ presents the form
\begin{align}
    \mathcal{E}(\rho(\bm{\theta})) &=
    \left[\bigcirc_{i=0}^{D-1}
    \left(\Lambda_{D - i} \circ \mathcal{U}_{D-i}\right)\right] \circ \nonumber\\
    &\quad \circ
    \left[\bigcirc_{j=1}^{D} \mathcal{U}_j^{-1}\right]
    \circ \rho(\bm{\theta})
    \label{eq: q channel e}
\end{align}
where \(\Lambda_l\) denotes the one-layer noise channel associated with the \(l\)-th application of the one-layer unitary transformation \(\mathcal{U}_l\). Fig. \ref{fig: circuit_paint}b shows this new setup. The error channel shown in this figure can be interpreted as going back from the ideal quantum state $\rho(\bm\theta)$ to the initialization step and obtaining it again, including the previously characterized noise from the quantum machine. Computing $\mathcal{E}(\rho)$ from Eq. (\ref{eq: q channel e}) by classical means is even more demanding than just computing the ideal output $\rho(\bm{\theta}) = \mathcal{U}_D \circ \cdots \circ \mathcal{U}_1 (\ket{0}\bra{0}^{\otimes n})$. However, if the noise level is sufficiently low, this formulation allows us to develop a more effective approach \cite{Filippov2023}, as we will discuss in Sec. \ref{sec: TN approach}.

During the execution of a noisy quantum circuit, multiple sources of error come into play. Among these, state preparation and measurement (SPAM) errors are particularly well-documented and represent a significant portion of the dynamic errors that arise during quantum computations \cite{Sun2018, Yu2023}. However, to simplify our analysis and focus on the core aspects of noise mitigation, we assume perfect initialization and measurement of the quantum state, setting aside the SPAM error problem, which has already been discussed in multiple studies \cite{Jayakumar2024, Lin2021, Geller2021, cai2023}. This assumption allows us to concentrate on understanding and mitigating the dynamic errors that occur during the application of quantum gates, which are the primary focus of our study. 

Typically, two-qubit quantum gates exhibit significantly higher noise levels compared to single-qubit rotations \cite{McCourt2023}. Bearing this in mind, we adopt the widely accepted sparse Pauli-Lindblad noise model \cite{ewout2023} to characterize and study the behavior of noise when executing two-qubit gates on the system. This noise model has recently gained attention in the QEM community due to its effective characterization and reliable results \cite{ewout2023, Filippov2023, vandenBerg2024, jaloveckas2023, Kim2020}.

\subsection{Tensor Network Approach}
\label{sec: TN approach}
In the rest of the paper, we adopt the Pauli Transfer Matrix (PTM) representation of quantum states, gates and operators. A comprehensive description of this representation can be found in Appendix~\ref{app: PTM}. The PTM formalism allows us to represent in compact form any linear operation $\mathcal{U} \in \mathcal{L}(\mathcal{H}, \mathcal{H})$ as a matrix $ B \in \mathbb{R}^{4^n} \times \mathbb{R}^{4^n}$, and quantum operators, such as density matrices $\rho \in \mathcal{D}(\mathcal{H})$ and squared observables $O \in \mathcal{L}(\mathcal{H})$, as a vector $V \in \mathbb{R}^{4^n}$.

Furthermore, we focus on linear topologies. That is, quantum systems characterized by a one-dimensional arrangement of qubits whose interactions are restricted to nearest-neighbor pairs. This decision is driven by the physical constraints of current hardware; most state-of-the-art quantum processors, particularly those based on superconducting circuits or trapped ions, face significant connectivity limitations due to the high overhead of cross-talk and the complexity of routing control lines in a multi-dimensional lattice \cite{Petersen2016, Abughanem2024}. By restricting the architecture to a 1D geometry, we can exploit the local nature of these interactions to model quantum impairments in real circuits more efficiently. 

Each unitary layer $\mathcal{U}_l$ is a set of quantum gates $U$ that is executed in the quantum system in parallel (i.e. at the same time). $\mathcal{U}_l$ can be represented by a matrix of dimension $4^n \times 4^n$, which poses significant challenges in terms of storage and computational operations due to its exponential size. However, by leveraging TN, its computational weight, including the cost of matrix multiplications, can be significantly reduced.

\begin{figure}[H]
    \centering
    \includegraphics[width=1.0\linewidth]{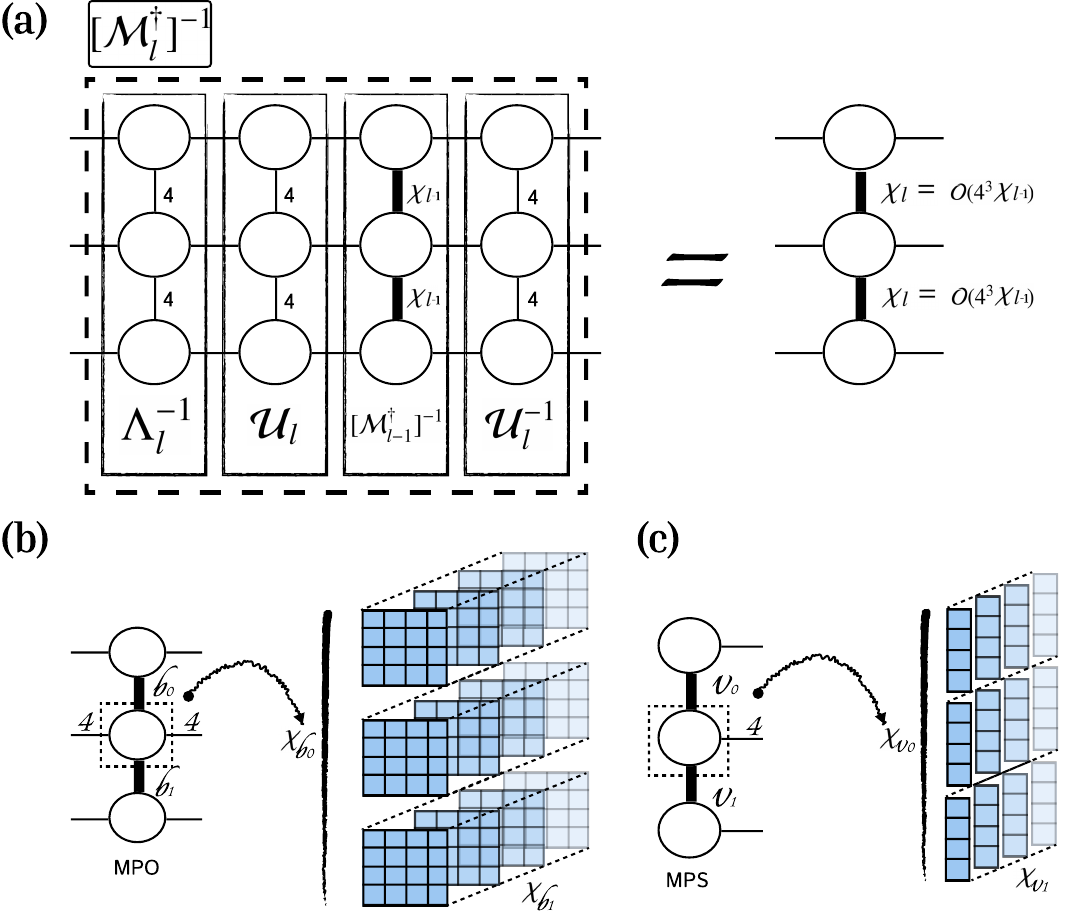}
    \caption{TN diagrams. (a) Middle-out contraction matrix $[\mathcal{M}_l^\dagger]^{-1}$ before (left) and after (right) TN contraction and compression. (b) Visual description of an MPO (left) and an MPS (right) of a 3-qubit system.} 
    \label{fig: TN diagram}
\end{figure}

For such one-dimensional systems, the Matrix Product Operator (MPO) formalism provides a powerful framework to describe the action of a quantum channel \cite{Orus2014}. This approach allows us to decompose global operations into a sequence of local tensors, effectively managing the potential growth of entanglement and ensuring that the simulation of noise and gate operations remains computationally tractable as the number of qubits grows. The MPO representation of any linear operator $B \in \mathbb{R}^{4^n} \times \mathbb{R}^{4^n}$ has the form of Eq. (\ref{eq: B MPO}), where, following the PTM formalism, $\mathcal{B}_{b_{l-1}, b_{l}}^{[l]} \in \mathbb{R}^{4} \times \mathbb{R}^{4}$ represents middle tensors of dimension 2 (i.e. a matrix) with virtual indices $b_{l-1}$ and $b_{l}$ acting on the $l$-th qubit. The bond dimension $\chi_b$ represents the dimension along each $b_l$ axis, with $l = 0, 1, ..., n-2$. Notice that the extreme elements in Eq. (\ref{eq: B MPO}) are special cases, with $\mathcal{B}_{b_0}^{[0]}$ and $\mathcal{B}_{b_{n-2}}^{[n-1]}$ being matrices with just one virtual index. In linear topologies, such TN representation reduces the memory cost of storage from an exponential scaling with $n$ to linear, $\mathcal{O}(4^2\chi^2 n)$. See Fig. \ref{fig: TN diagram}b for a visual representation of an MPO.

Any quantum operator, such as density matrices and observables, can also be efficiently represented using TN in such linear topologies. In this case, their vector representation \( V \in \mathbb{R}^{4^n} \) can be mapped to a Matrix Product State (MPS) \cite{Orus2014} following Eq. (\ref{eq: V MPO}), where, considering the PTM formalism, $\mathcal{V}_{v_{l-1}, v_{l}}^{[l]} \in \mathbb{R}^4$ represents middle tensors of dimension 1 (i.e. a vector) with virtual indices $v_{l-1}$ and $v_{l}$ acting on the $l$-th qubit. The bond dimension $\chi_v$ represents the dimension along each $v_l$ axis. Notice that the edge tensors are special cases, with $\mathcal{V}_{v_0}^{[0]}$ and $\mathcal{V}_{v_{n-2}}^{[n-1]}$ being vectors with just one virtual index. See Fig. \ref{fig: TN diagram}c for a visual representation of an MPS.

As shown in Ref. \cite{Filippov2023}, for low noise levels, the MPO structure of $\mathcal{E}(\sbullet)$ becomes computationally efficient when exploiting the fact that each layer $\mathcal{U}_l^{-1}$ approximately cancels with its corresponding noisy map $\mathcal{U}_l \circ \Lambda_l$, which ensures a TN representation with low bond dimension.
\vspace{0.6cm}

As illustrated in Fig. \ref{fig: circuit_paint}b, the most efficient MPO $\mathcal{M}$ that captures the multi-layer error map $\mathcal{E}(\sbullet)$ is constructed by implementing the contractions from the middle (where the noisy circuit ends and the ideal inverted circuit starts) and proceeding outwards. As shown in Fig. \ref{fig: circuit_paint}b, the $l$-th iteration contraction has the form
\begin{equation}
    \mathcal{M}_l = \Lambda_l \circ \mathcal{U}_l \circ \mathcal{M}_{l-1} \circ \mathcal{U}_l^{-1},
\end{equation}
which involves two layers on the left side and one layer on the right side.

In order to express the noise $\Lambda_l$ as an MPO, one needs to efficiently characterize the noise associated with its corresponding unitary operation $\mathcal{U}_l$. Currently, there exist several state-of-the-art noise characterization techniques, such as the topography of individual noisy gates \cite{Nielsen2021}. However, as we have previously stated, we focus on the Pauli-Lindblad model with nearest-neighbor crosstalk \cite{ewout2023, Berg2024}, where noise can be characterized through a polynomial number of fidelities for
\begin{widetext}
\begin{equation}
    B = \sum_{b_0, ..., b_{n-2} = 0}^{\chi_b -1} \mathcal{B}_{b_0}^{[0]} \otimes \mathcal{B}_{b_0, b_1}^{[1]} \otimes \cdots \otimes \mathcal{B}_{b_{n-3}, b_{n-2}}^{[n-2]} \otimes \mathcal{B}_{b_{n-2}}^{[n-1]},
    \label{eq: B MPO}
\end{equation}
\begin{equation}
    V = \sum_{v_0, ..., v_{n-2} = 0}^{\chi_v -1} \mathcal{V}_{v_0}^{[0]} \otimes \mathcal{V}_{v_0, v_1}^{[1]} \otimes \cdots \otimes \mathcal{V}_{v_{n-3}, v_{n-2}}^{[n-2]} \otimes \mathcal{V}_{v_{n-2}}^{[n-1]},
    \label{eq: V MPO}
\end{equation}
\end{widetext}
some Pauli terms. Each noisy layer $\Lambda_l$ can be represented as a sequence of commuting two-qubit Pauli channels applied to adjacent qubits, which results in an MPO of bond dimension $\chi_{\Lambda} = 4$.

Since each ideal unitary layer $\mathcal{U}$ has at most bond dimension $\chi_u = 4$ (as it is composed only of single-qubit rotations and two-qubit gates), the bond dimension of the middle-out contraction $\mathcal{M}_l$ scales as
\begin{equation}
    \chi_l = 4^{3}\chi_{l-1}
\end{equation}
where $\chi_{l-1}$ is the bond dimension of the previous iteration $\mathcal{M}_{l-1}$. For a circuit of depth $L$, this results in an exponentially growing bond dimension of $\mathcal{O}(4^{3L})$, which is untractable in practice for large $n$. To overcome this problem, the MPO \(\mathcal{M}_l\) can be compressed after each iteration either to a fixed maximal bond dimension \(\chi_\text{max}\) or to a desired precision. The latter is achieved by truncating the smallest singular values in the canonical representation of the MPO or by employing variational methods \cite{Hubig2017}. In our work, we apply the Randomized Singular Value Decomposition (RSVD) \cite{Halko2011} to perform the compression to a fixed bond dimension, as it has proven to be computationally more efficient than its direct SVD counterpart. Each time a compressed MPO $\mathcal{M}$ is obtained, we are actually performing an efficient approximation of the multi-layer map $\mathcal{E}(\sbullet)$. The computational cost of MPO compression scales as \(\mathcal{O}(n \chi^3)\) \cite{Hubig2017}, which significantly exceeds the cost of MPO multiplication $\mathcal{O}(n \chi^2)$, so we spend more resources on the compression of the MPO structure rather than in their contractions.

\subsection{Surrogate observable}
\label{sec: surrogate}
Our goal is to find the surrogate observable:
\begin{equation}
    \hat{Y} \quad \text{s.t} \quad \mathcal{E}^\dagger(\hat{Y}) = X \Leftrightarrow \left[\mathcal{E}^\dagger\right]^{-1} (X) = \hat{Y}.
    \label{eq: objective}
\end{equation}

The exact solution for $\hat{Y}$ is given by inverting the whole adjoint channel, $\left[\mathcal{E}^\dagger\right]^{-1}$. Generally, not all CPTP maps are invertible \cite{Nielsen2000}. However, the considered Pauli-Lindblad model allows for an inversion \cite{ewout2023}. Taking into account Eq. (\ref{eq: q channel e}), the inverse map is given by

\begin{align}
    \left[\mathcal{E}^\dagger\right]^{-1}(\sbullet) &=
    \left[\bigcirc_{i=0}^{D-1}
    \left(\Lambda_{D - i}^{-1} \circ \mathcal{U}_{D-i}\right)\right] \circ \nonumber\\
    &\quad \circ
    \left[\bigcirc_{j=1}^{D} \mathcal{U}_j^{-1}\right]
    \label{eq: middle out one}
\end{align}
where we used $\mathcal{U}^{-1} = \mathcal{U}^\dagger$, $(AB)^\dagger = B^\dagger A^\dagger$, $(AB)^{-1} = B^{-1}A^{-1}$, $\Lambda^\dagger = \Lambda$, $(\Lambda^{-1})^\dagger = \Lambda^{-1}$ as the Pauli-Lindblad noise is diagonal in the PTM representation. Thus, we can efficiently construct the MPO $[\mathcal{M}^\dagger]^{-1}$ that represents the inverse adjoint map $\left[\mathcal{E}^\dagger\right]^{-1}(\sbullet)$ by iteratively applying tensor contractions. The $l$-th iteration of such a middle-out contraction has the form
\begin{equation}
    [\mathcal{M}^\dagger]_l^{-1} = \Lambda_l^{-1} \circ \mathcal{U}_l \circ [\mathcal{M}^\dagger]_{l-1}^{-1} \circ \mathcal{U}_l^{-1}.
    \label{eq: middle out inverse}
\end{equation}
See Fig. \ref{fig: TN diagram}a for a schematic depiction. Notice that, while such recursive iteration coincides with the one described in TEM~\cite{Filippov2023}, the underlying approach to the problem is fundamentally different. TEM reconstructs the noisy quantum state by using a set of Dual Operators $\{D_i\}_i$, obtained through the action of an Informationally Complete Positive Operator-Valued Measurements (IC-POVM). On the contrary, we evolve the target observable through the Heisenberg formalism, in order to obtain a surrogate observable whose expectation value matches that of the target observable.   

Being a valid observable, the surrogate $\hat{Y}$ can be expressed in the Pauli basis. From Eq. (\ref{eq: objective}), (\ref{eq: middle out one}), (\ref{eq: middle out inverse}) we have
\begin{equation}
    \hat{Y} = \sum_{i=0}^{4^n-1} y_i P_i,
\end{equation}
where $\bm{y} = [\mathcal{M}^\dagger_L]^{-1} \cdot \bm{x}$. This means that whenever the target observable is a Pauli, that is $X = P_i$, then $\hat{Y} = [\mathcal{M}_L^\dagger]^{-1} P_i = [\mathcal{M}_L^\dagger]^{-1}_{:, i} $ is the $i$-th column of the middle-out contraction matrix (i.e. recall that $P_i$ has a vector form in the PTM formalism). Therefore, we have 

\begin{equation}
    \left< \hat{Y} \right>_{\mathcal{E}(\rho)} = \sum_{k=0}^{4^n-1} [\mathcal{M}_L^\dagger]^{-1}_{k,i} \left< P_k \right>_{\mathcal{E}(\rho)},
\end{equation}
where $[\mathcal{M}_L^\dagger]^{-1}_{k,i}$ is the element associated to the $i$-th column and $k$-th row of the middle-out contraction matrix (see Eq. (\ref{eq: middle out inverse})) at the $L$-th iteration, and $\left< P_k \right>_{\mathcal{E}(\rho)}$ is the noisy expectation value of Pauli string $P_k$ measured over the noisy quantum state $\mathcal{E}(\rho)$ that the quantum hardware outputs. 

If no tensor compression is performed to truncate the bond dimension, we should expect to fulfill Eq. (\ref{eq: final objective}). Nevertheless, compressing the MPO will inevitably induce errors that cannot be avoided. We assess the validity of such errors numerically in Section \ref{sec: results}.

\section{Measurement Overhead}
The measurement overhead $\gamma$ is a metric for evaluating and comparing the efficiency of QEM protocols. It is defined as the ratio of the standard deviation of an estimated observable after applying a noise mitigation protocol $\Delta O_\text{e.m.}$ with respect to the standard deviation before the mitigation $\Delta O_\text{noisy}$. That is:
\begin{equation}
    \gamma = \frac{\Delta O_\text{e.m.}}{\Delta O_\text{noisy}}.
    \label{eq: sample over}
\end{equation}
This metric provides insights into the additional resources required to achieve reliable quantum computations in the presence of noise. Specifically, the squared value of the overhead $\gamma^2$ directly quantifies the factor number of circuit repetitions (shots) needed to attain the desired precision, making it a key indicator of the practical feasibility of QEM techniques \cite{Filippov2023, ewout2023, Takagi2022, Takagi2023, Tsubouchi2023, Xiong2020, Xiong2022, Hsieh2024}.

Recent works, such as those by Tsubouchi \textit{et al.} \cite{Tsubouchi2023} and Takagi \textit{et al.} \cite{Takagi2022, Takagi2023} showed that the measurement overhead must grow exponentially with the depth of the quantum circuit, highlighting a fundamental limitation in the scalability of current QEM methods. Among the various QEM protocols developed to date, the TEM technique \cite{Filippov2023} stands out as the most efficient in terms of measurement overhead. TEM has been proven to achieve the theoretical lower bound for overhead scaling, making it the optimal choice for minimizing the number of shots. 

In this study, we employ TN methods to model noise within the framework of the Pauli-Lindblad model \cite{ewout2023}, mirroring the setup used to evaluate TEM \cite{Filippov2023} for the sake of pairness. Representing noise dynamics through tensor networks yields an efficient method to calculate the surrogate observable using tensor contractions implemented on classical hardware. As we show in Sec. \ref{sec: benchmarking}, our methodology approaches the theoretical lower bound \cite{Filippov2023}, slightly improving over TEM.

\begin{figure}
    \centering
    \includegraphics[width=1.0\linewidth]{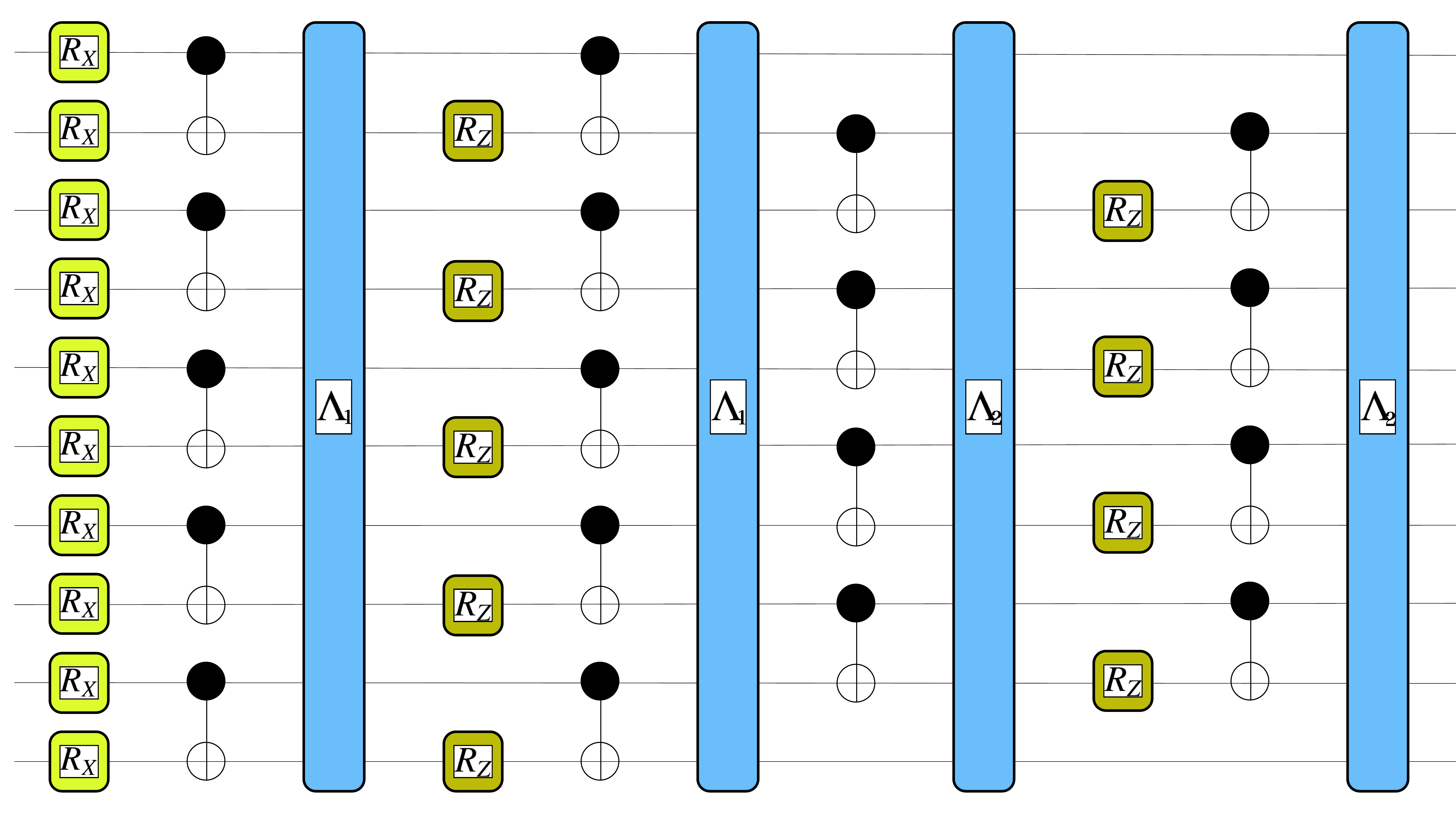}
    \caption{\footnotesize{A single Trotter step for the one-dimensional transverse-field Ising model. First layer consists of a set of unitary rotations $R_X(2h\delta_t)$ for every qubit and two distinct implementations of pairwise $ZZ$-rotations (for even and odd links between the qubits), each consisting of two repeated CNOT layers intervened by the unitary rotation $R_Z(-2J\delta_t)$ on controlled qubits. Model parameters are $h = 1$, $J = 0.5236$, and $\delta_t = 0.5$, reproducing the settings of TEM \cite{Filippov2023}. Each unique CNOT layer is followed by a sparse Pauli-Lindblad noise \cite{ewout2023}, denoted as $\Lambda_1$ and $\Lambda_2$, with sampling overhead $\gamma_1 = 1.140$ and $\gamma_2 = 1.137$, respectively.}}
    \label{fig: one trotter step}
\end{figure}

\section{Results}
\label{sec: results}
In order to compare our methodology with PEC \cite{ewout2023} and TEM \cite{Filippov2023}, we consider the discrete-time 10-qubit dynamics of the one-dimensional transverse-field Ising model. The diagram in Fig. \ref{fig: one trotter step} shows the model for one Trotter step. A detailed description of such a model is provided in the caption. We assume that the noisiness of single-qubit gates is negligible when compared to two-qubit gates, as confirmed by reasurements on simple local rotations \cite{Noiri2022, Aseguinolaza2024}, and that the noise produced during each layer of CNOT gates can be described by the widely adopted sparse Pauli-Lindblad model \cite{ewout2023}. The model parameters, including noise rates, are the same as in Ref. \cite{Filippov2023}, which ensures a consistent and fair comparison. TN contractions and compressions are implemented using the Quimb package \cite{Gray2018}, while  Qibo \cite{Efthymiou2021} is employed for quantum circuit simulations.

\subsection{Approximation of the evolved observable}
\label{sec: Structure Matrix}
The exact evaluation of $\hat{Y}$ would require measuring all Pauli components $\{P_i\}_{i=0}^{4^n}$ and rescaling each outcome by its associated value $\{y_i\}_{i=0}^{4^n}$. This direct approach is unfeasible, since it implies an exponential number of measurements on the quantum circuit. An alternative would be to measure $\hat{Y}$ through probabilistic sampling, as in PEC. That is, we can sample and measure a Pauli string $P_i$ according to some probability distribution $p_i$, and estimate the expectation value as
\begin{equation}
    \left<\hat{Y} \right> = \gamma \sum_{i=0}^{4^n-1} \text{sgn}(y_i)p_i \left<P_i\right>,
\end{equation}
with $\text{sgn}(\cdot)$ being the sign function, $p_i = \frac{|y_i|}{\gamma}$ and $\gamma = \sum_{i=0}^{4^n-1} |y_i|$. It can be proven that $\gamma$ corresponds to the importance sampling of the probabilistic evaluation of the observable \cite{ewout2023}. In practice, $\gamma$ grows fast with the depth of the circuit, as an exponential number of terms are considered in the sum, which directly impacts the variance of the expectation value estimator. This problem renders the probabilistic approach impractical.

By examining the data from extensive numerical simulations, we observed that, for the considered setup, the middle-out contraction MPO $[\mathcal{M}_L^\dagger]^{-1}$ has a matrix representation that keeps a nearly diagonal shape during all Trotter steps. Such a structure is not due to TN compression and also appears when exact matrix multiplication (i.e, without compression) is performed.

\begin{figure}
    \centering
    \includegraphics[width=1.0\linewidth]{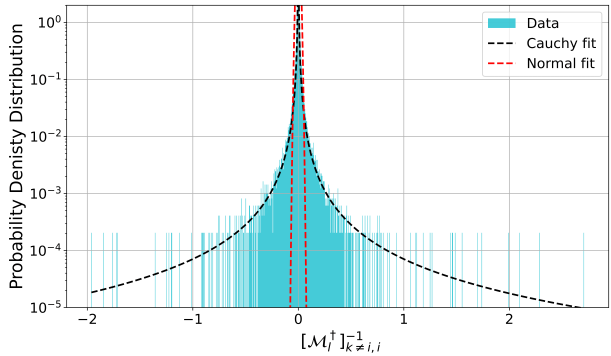}
    \caption{Histogram representing the probability density function of the off-diagonal matrix elements $\big \{[\mathcal{M}_l^\dagger]^{-1}_{k\neq i,i} \big\}_k$ when $P_i = Z^{\otimes 10}$ at step 18 of the Trotterization of the Ising model. The histogram is computed using fixed-width bins and normalized such that the total area under the distribution equals one. The resulting density estimate is then compared against a Cauchy (black dots) and a Normal (red dots) parametric fit. Fitting was performed by the SciPy Python library \cite{2020SciPy-NMeth}.}
    \label{fig: histogram step 18}
\end{figure}

As an example, let us consider $P_i = Z^{\otimes 10}$. Fig. \ref{fig: histogram step 18} shows the distribution of the values $\big \{[\mathcal{M}_l^\dagger]^{-1}_{k\neq i,i}\big\}_k$ at trotter step 18. The largest off-diagonal value ($k \neq i$) in Fig. \ref{fig: histogram step 18} is on the order of $\sim 1$, while the diagonal term $ [\mathcal{M}_l^\dagger]^{-1}_{i, i}$ (not shown in the picture) is on the order of $\sim 10^2$. As can be seen in Fig. \ref{fig: histogram step 18}, the Cauchy distribution is, among those considered, the one that best fits the data. 

Since $[\mathcal{M}_l^\dagger]^{-1}_{i,i} \gg [\mathcal{M}_l^\dagger]^{-1}_{k,i}, \forall k \neq i$, and $\left< P_k \right> \in [-1,1]$, $\hat{Y}$ can be approximated by the target initial Pauli $P_i$ scaled by the diagonal matrix element $[\mathcal{M}_l^\dagger]^{-1}_{i,i}$, that is:
\begin{equation}
    \hat{Y} \approx [\mathcal{M}_L^\dagger]^{-1}_{i,i} P_i,
    \label{eq: approx Y}
\end{equation}
which translates to an expectation value of
\begin{equation}
    \left< \hat{Y} \right>_{\mathcal{E}(\rho)} \approx [\mathcal{M}_L^\dagger]^{-1}_{i,i} \left< P_i \right>_{\mathcal{E}(\rho)}.
    \label{eq: final Y}
\end{equation}
We refer to this first-order approximation as Dominant Component Approximation (DCA). As we will discuss in section \ref{sec: bias}, this approximation yields very accurate results even for deep circuits, achieving a comparatively low bias and a tight sampling overhead.

\begin{figure*}
\centering
    \includegraphics[width=1\linewidth]{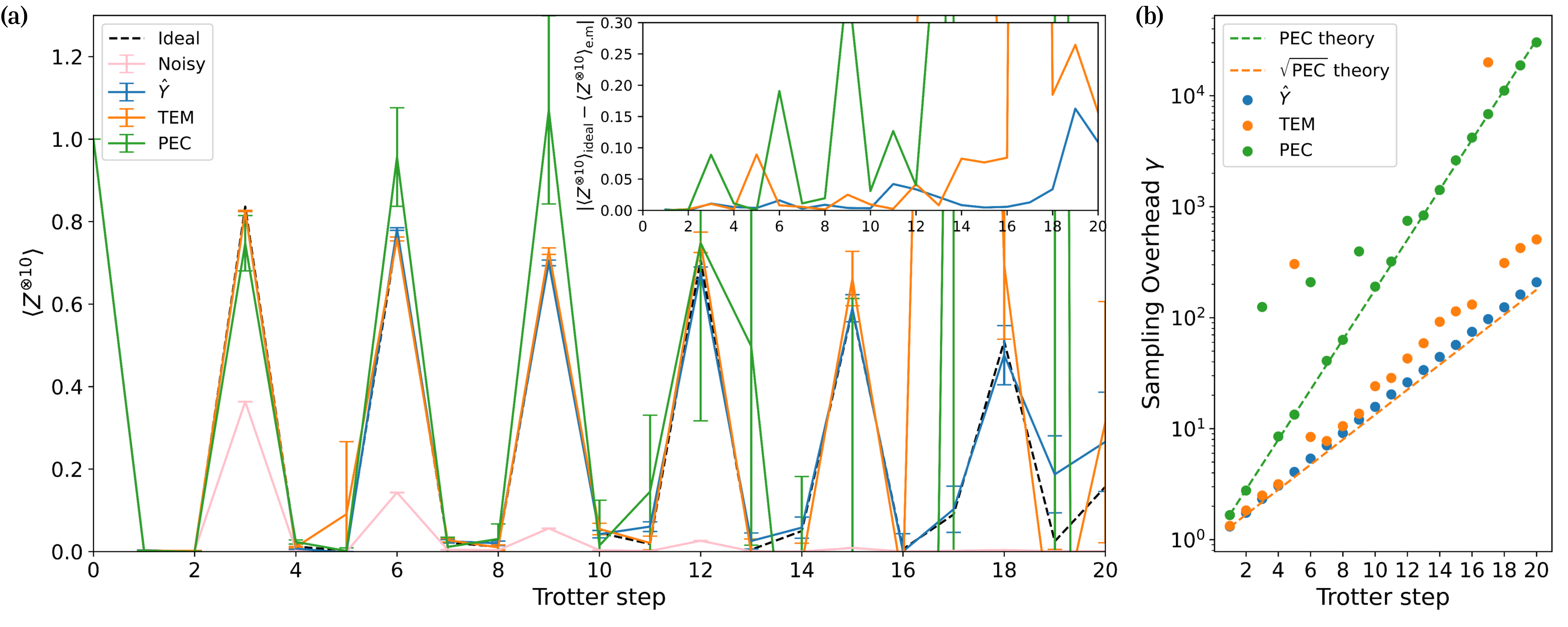}
    \caption{Comparison of PEC, TEM, and the mitigation procedure proposed in this article with $\hat{Y}$. We mitigate the expectation value $\left<Z^{\otimes 10}\right>$ for a 10-qubit discrete-time evolution of the Ising model. (a) Dynamics of the expectation value $\left<Z^{\otimes 10}\right>$ with and without noise mitigation. PEC estimation is based on 300 circuits sampled from the quasi-probability representation of the inverse noise, with 10000 shots per circuit on the computational basis. TEM estimation is based on 300 circuits, with projective measurements in local Paulis $\sigma_1$, $\sigma_2$ or $\sigma_3$ for every qubit, and 10000 shots per circuit; these measurements basis are chosen with probabilities $p_1=0.001$, $p_2=0.001$ and $p_3=0.998$ to adjust for this concrete observable; mitigation using $\hat{Y}$ is performed in one circuit with a total of $3\times10^6$ shots. The bond dimension of the noise mitigation map is at most 200. (b) Sampling overhead in the numerical experiments (dots) estimated as the ratio of the noise-mitigated and unmitigated estimation errors, $\gamma = \Delta \hat{O}_\text{n.m.} / \Delta \hat{O}_\text{noisy}$, as well as their theoretical predictions (dashed lines).}
    \label{fig: exp_val}
\end{figure*}

\subsection{Benchmarking}
\label{sec: benchmarking}
In order to compare all methods on equal footing, we keep the total number of shots (i.e., the number of circuit runs) constant. For PEC and TEM, we use 300 different circuits with $10^4$ shots each. PEC requires sampling unitary gates from the quasi-probability distribution of the inverse noise map. TEM requires implementing an informationally complete positive operator-valued measure (IC-POVM) \cite{scott2004, Filippov2023}, performed through sampling projective measurements on a different basis to capture the structure of the noisy quantum state $\mathcal{E}(\rho(\bm{\theta}))$ that the circuit outputs. This allows us to construct the so-called Dual Operators \cite{Filippov2023} in a TN structure. Dual Operators will be used in post-processing TN contractions to estimate the expectation value of the observable $\left<O \right>$ and its standard deviation $\Delta O$. As for our method, we directly measure the noisy circuit using the observable proposed in Eq. (\ref{eq: approx Y}), with a total of $3\times10^6$ measurement shots. As shown in Eq. (\ref{eq: final Y}), the noisy measurement outcomes are then rescaled by a parameter. As in TEM, we choose the target $X = Z^{\otimes 10}$, which corresponds to standard measurements on the computational basis.

For the considered noise model, if no mitigation strategies are in place, the noisy estimation of the observable $Z^{\otimes 10}$ quickly decays and deviates from the true value, as shown in Fig. \ref{fig: exp_val}a. We appreciate how TN techniques surpass the PEC strategy. Notice that both TEM and the proposed DCA estimator $\hat{Y}$ start to diverge at the same Trotter step. Both are implemented using the same TN strategy to capture the noise in the circuit, and our methodology performs similarly to TEM, slightly improving it after Trotter step 14. This indicates that our approximation, despite its simplicity and direct construction, successfully captures the essential features of the noise model.

Let us now compare the sampling overhead $\gamma$ for the different methods. The values of $\gamma$ are depicted in Fig. \ref{fig: exp_val}b and are computed using Eq. (\ref{eq: sample over}). The PEC protocol follows the expected theoretical tendency $\gamma_{\text{PEC}}$ \cite{ewout2023}, and TEM is close to the theoretical minimum at low circuit depths, but deviates from it as the depth of the circuit increases. Our DCA method closely approaches the theoretical lower bound $\gamma_{\hat{Y}} \approx \sqrt{\gamma_{\text{PEC}}}$ at all considered circuit depths, improving over both PEC and TEM. It is also worth mentioning that these two methodologies occasionally show a higher sampling overhead in some specific steps, distancing themselves from the theoretical tendency. On the contrary, our approach exhibits reduced variability and a tighter alignment with the expected trend. This results in a more stable and predictable pattern, with fewer outliers and a closer fit to the theoretical exponential behavior. As a last remark, our DCA consists in scaling the unmitigated outcomes by the diagonal element $[\mathcal{M}_L^\dagger]^{-1}_{i,i}$ (see Eq. (\ref{eq: final Y})). For for this reason, the exponential scaling of the sampling overhead is simply $\gamma_{\hat{Y}} = [\mathcal{M}_L^\dagger]^{-1}_{i,i}$.

\section{Discussion}
Although fundamentally different, our method shares with TEM \cite{Filippov2023} the fact that both require the truncation of the bond dimension for the MPO $[\mathcal{M}_L^\dagger]^{-1}$. In this section, we estimate the expectation value of the surrogate $\hat{Y}$ for a set of bond values $\chi$ considered during MPO and MPS contraction. 


\subsection{Bias of the surrogate}
\label{sec: bias}
\begin{figure}[H]
    \centering
    \includegraphics[width=1.0\linewidth]{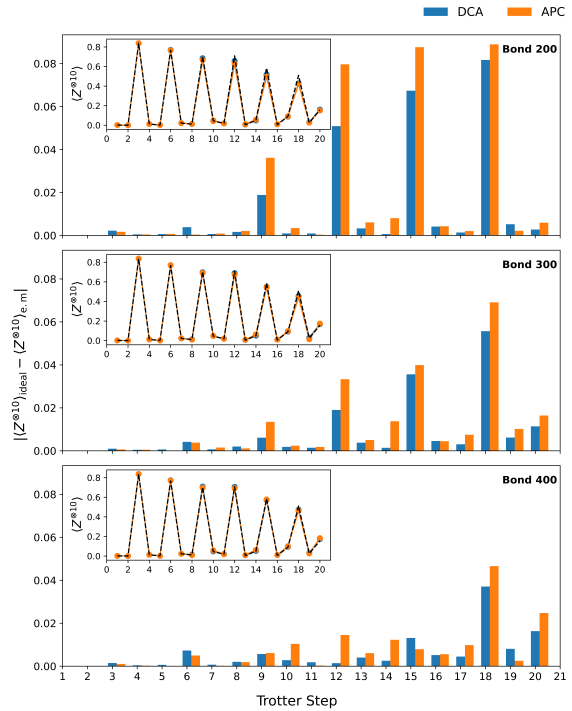}
    \caption{Absolute difference between the ideal and noise-mitigated outcome $\big |\left<Z^{\otimes 10}\right>_\text{ideal} - \left<Z^{\otimes 10}\right>_\text{e.m.} \big |$ as a function of the trotter step, considering the Dominant Component Approximation (blue) and all Pauli components (orange). Different bond dimension truncations were considered, namely: 200, 300, and 400. The inset indicates the mitigated outcomes compared to the ideal one (dashed lines) as a function of the Trotter step.}
    \label{fig: comparison 3}
\end{figure}

The DCA involves discarding an exponentially large number of terms, which should introduce bias in our simplified estimator. For this reason, we proceed to examine the expectation value error produced by the implementation of the DCA protocol, and compare it with the case in which all Pauli components (APC) are taken into account. Fig. \ref{fig: comparison 3} shows the absolute error of the mitigation strategy for DCA and APC. The results presented here are based on numerical simulations of noisy quantum circuits carried out with the Qibo simulator.

At low circuit depths (steps 1 to 8), the difference between the DCA and the exact solution obtained with APC is negligible. At such depths, using all off-diagonal terms $\big \{[\mathcal{M}_L^\dagger]^{-1}_{k\neq i,i} \big\}_k$ does not seem to improve over the DCA. At higher depths (beyond step 8), we observe a tendency for the DCA to exhibit reduced errors compared to the full APC solution. This is counterintuitive, as one would expect to get a more accurate representation of $\hat{Y}$ when including off-diagonal terms. Increasing the bond dimension up to 400, we observe a direct and positive impact on the mitigation process, reducing the error in both the DCA and APC methodologies. However, the overall trend observed for bond dimension 200 remains consistent: including all off-diagonal Pauli terms still leads to a degradation of the mitigated outcome at larger circuit depths, suggesting that these contributions may effectively behave as noise. This effect likely arises from an insufficient bond dimension to capture all relevant information at such depths. Increasing the bond dimension will eventually reduce the error and render all Pauli terms meaningful at some bond value $\chi$, but it remains unclear how large the bond must be for this to occur. In the example at hand, doubling the bond dimension from 200 to 400 does not lead to a significant improvement in the APC performance over the DCA at high depths. Since DCA also benefits from larger bond dimensions, achieving the bond size required for APC to surpass the DCA could lead to intractably large bond dimensions for classical systems to deal with.

\subsection{No classical description of the (noisy) quantum state}
The main computational advantage of our approach compared to TEM \cite{Filippov2023} is that it does not require the construction of the Dual Operators \cite{Innocenti2023}. Apart from its complexity, this process often presents significant challenges related to the identification of an IC-POVM capable of extracting sufficient information from the quantum circuit’s output. Recent studies highlight the non-triviality of this task, with multiple works proposing specialized optimization techniques to derive POVMs that yield the optimal Dual Operators \cite{Fischer2024, Malmi2024}.

Besides, not all combinations of quantum systems and observables allow for an efficient construction of Dual Operators from single-qubit measurements. The formation of global duals, which cannot be decomposed as tensor products of local terms $\bigotimes D_i$, often exhibits superior accuracy in capturing correlations, but their TN representation manifests higher bond dimension $\chi$. This, in turn, increases the complexity of the TN contractions performed in TEM. As an example, in our simulations, TEM requires constructing as many Dual Operators as measurements are performed, each one expressed as an MPS of bond dimension $\chi_D=1$. Then, each Dual Operator MPS is contracted with the middle-out contraction MPO $[\mathcal{M}_L^\dagger]^{-1}$ and the target observable MPS $X$. This implies that $3\times10^6$ tensor contractions need to be performed in our example. Increasing the bond dimension of the Dual Operator MPS  into $\chi_D$ results in an increase in the complexity of each single tensor contraction, which increases the overall computational time of the protocol as $\mathcal{O}(\chi_D^2)$. Our surrogate observable approach does not suffer from this bottleneck since it does not require the construction of Dual Operators.

\subsection{Computational complexity}
One of the main advantages of the protocol presented in this paper with respect to TEM lies in its computational efficiency. Consider a Pauli observable $O = P_i$. In this specific case, TEM's classical complexity for the whole set of TN contractions scales roughly as 

\begin{equation}
    \mathcal{C}_\text{TEM} = \mathcal{O}(Mnd\chi^2),
\end{equation} 
where $n$ is the number of qubits, $d=4$ is the physical index (which spans the Hilbert space in the PTM representations), $\chi$ is the bond dimension, and $M$ represents the number of contractions needed for each set of Dual Operators, which corresponds to the number of circuit executions. $M$ is usually treated as a tunable parameter, as it determines the precision $\epsilon$ of the expected value achieved by measuring the quantum circuit in a given observable, which scales as $\epsilon \sim \mathcal{O}(1/\sqrt{M})$. Practical scenarios require $M \sim 10^6$ circuit executions, which makes it a significant factor that cannot be ignored when discussing the complexity of the procedure. 

With this respect, the proposed DCA protocol greatly reduces the computational complexity, as it only requires measuring the diagonal element of the middle-out contraction matrix. Consequently, a single TN contraction is required for a given Pauli observable. This results in a computational cost that scales as

\begin{equation}
    \mathcal{C}_\text{DCA} = \mathcal{O}(nd\chi^2).
\end{equation} 
The ratio of the complexities in the two methods is, then:

\begin{equation}
    \frac{\mathcal{C}_\text{TEM}}{\mathcal{C}_\text{Y}} \sim M.
\end{equation}

In other words, the proposed DCA is $M$ times (i.e., the number of circuit executions) more efficient than TEM. In practical scenarios, this might imply a speed-up advantage of $\sim 10^6$ fewer TN contractions. This improvement was observed in our simulations with a maximum bond of 200: leaving aside the computation of the MPO $[\mathcal{M}^\dagger_L]^{-1}$ (as both the DCA and TEM use the same TN procedure), the mitigation for the entire 20-step evolution under the Ising Hamiltonian took 10 days to complete for the TEM strategy, whereas our approach achieved the same result in less than 5 seconds, both performed on the same CPU (Intel Xeon $2.5$GHz) with 16 cores and 100 GB of RAM.

\section{Conclusion}
In this work, we addressed noise mitigation of NISQ hardware by constructing a surrogate observable \( Y \), whose expectation value over the noisy quantum state \(\mathcal{E}(\rho)\) reproduces the same result as the expectation value of a target observable \( X \) over the noise-free state \(\rho\). The main obstacle is its practical implementation, which became infeasible due to the exponential growth of complexity with the system size. In particular, standard tomographic techniques for characterizing the effect of the noisy quantum channel, as well as direct matrix multiplications involving high-dimensional operators, are computationally intractable for large systems. To overcome this issue, we applied the middle-out TN contraction methodology. By using the TN representation, it is possible to efficiently compute a good approximation to the surrogate observable $\hat{Y}$ that efficiently mitigates the error presented in a noisy quantum computer with a sparse Pauli-Lindblad noise description, and that achieves the QCRB for unbiased estimators. Specifically, when the target observable is a Pauli string $P_i$, $\hat{Y}$ can be effectively approximated by a rescaled version of $P_i$. Such a low-complexity approach shows lower bias than TEM at deeper circuits, while significantly reducing the computational complexity with respect to state-of-the-art techniques, as just a single TN contraction needs to be performed.

As a concluding remark, recent works have identified fundamental limitations that no unbiased QEM strategy can surpass. Nonetheless, the practical implementation of QEC remains severely constrained by current quantum hardware, rendering QEM the primary viable approach for noise mitigation in near-term quantum devices. Although QEM may appear to be a transient research direction, destined to become obsolete once fully fault-tolerant quantum computation is achieved through QEC protocols, this view is increasingly being challenged. Rather than competing paradigms, QEC and QEM should be understood as complementary methodologies whose combined application can substantially enhance quantum computational performance. Indeed, recent studies indicate that the integration of QEC and QEM constitutes one of the most promising avenues currently available \cite{Zhou2025}. Accordingly, QEM is unlikely to disappear with the advent of fault-tolerant quantum computers; instead, it is expected to play a pivotal role in accelerating the transition toward fault tolerance. In this context, it is therefore essential to develop novel QEM strategies that simultaneously reduce bias and variance while maintaining manageable computational complexity.

\section{Acknowledgments}
This work has been funded by grants PID2022-137099NB-C41 funded by MCIN/AEI/10.13039/501100011033 and FSE+ and by grant 2021 SGR 01033 funded by AGAUR, Dept. de Recerca i Universitats de la Generalitat de Catalunya
10.13039/501100002809. The work of G. Cocco is supported by the Ramon y Cajal fellowship program (grant RYC2021-033908-I) funded by the Spanish Ministry for Science and Innovation/State Research Agency MCIN/AEI/10.13039/501100011033 and by the European Union «NextGenerationEU» Recovery Plan for Europe.

\bibliographystyle{apsrev4-2-new}
\bibliography{references}

@article{Kivlichan2020,
   author = {Ian D. Kivlichan and Craig Gidney and Dominic W. Berry and Nathan Wiebe and Jarrod McClean and Wei Sun and Zhang Jiang and Nicholas Rubin and Austin Fowler and Alán Aspuru-Guzik and Hartmut Neven and Ryan Babbush},
   doi = {10.22331/q-2020-07-16-296},
   issn = {2521327X},
   journal = {Quantum},
   month = {7},
   pages = {296},
   publisher = {Verein zur Förderung des Open Access Publizierens in den Quantenwissenschaften},
   title = {Improved Fault-Tolerant Quantum Simulation of Condensed-Phase Correlated Electrons via Trotterization},
   volume = {4},
   url = {https://quantum-journal.org/papers/q-2020-07-16-296/},
   year = {2020},
}

@article{Lee2021,
   author = {Joonho Lee and Dominic W. Berry and Craig Gidney and William J. Huggins and Jarrod R. McClean and Nathan Wiebe and Ryan Babbush},
   doi = {https://doi.org/10.1103/PRXQuantum.2.030305},
   issn = {26913399},
   issue = {3},
   journal = {PRX Quantum},
   month = {9},
   pages = {030305},
   publisher = {American Physical Society},
   title = {Even More Efficient Quantum Computations of Chemistry through Tensor Hypercontraction},
   volume = {2},
   url = {https://doi.org/10.1103/PRXQuantum.2.030305},
   year = {2021},
}

@article{Acharya2024,
   author = {Rajeev Acharya and Dmitry A. Abanin and Laleh Aghababaie-Beni and Igor Aleiner and Trond I. Andersen and Markus Ansmann and Frank Arute and Kunal Arya and Abraham Asfaw and Nikita Astrakhantsev and Juan Atalaya and Ryan Babbush and Dave Bacon},
   doi = {10.1038/s41586-024-08449-y},
   issn = {1476-4687},
   journal = {Nature},
   month = {12},
   pages = {1-3},
   publisher = {Nature Publishing Group},
   title = {Quantum error correction below the surface code threshold},
   url = {https://www.nature.com/articles/s41586-024-08449-y},
   year = {2024},
}

@article{Campbell2024,
   author = {Earl Campbell},
   doi = {10.1038/s42254-024-00706-3},
   issn = {2522-5820},
   issue = {3},
   journal = {Nat. Rev. Phys.},
   month = {2},
   pages = {160-161},
   publisher = {Nature Publishing Group},
   title = {A series of fast-paced advances in Quantum Error Correction},
   volume = {6},
   url = {https://www.nature.com/articles/s42254-024-00706-3},
   year = {2024},
}

@article{Chatterjee2023,
   author = {Avimita Chatterjee and Koustubh Phalak and Swaroop Ghosh},
   title = {Quantum Error Correction For Dummies},
   journal = {arXiv:2304.08678},
   url = {http://arxiv.org/abs/2304.08678},
   year = {2023},
}

@article{Abobeih2022,
   author = {M. H. Abobeih and Y. Wang and J. Randall and S. J.H. Loenen and C. E. Bradley and M. Markham and D. J. Twitchen and B. M. Terhal and T. H. Taminiau},
   doi = {10.1038/s41586-022-04819-6},
   issn = {1476-4687},
   issue = {7916},
   journal = {Nature},
   month = {5},
   pages = {884-889},
   pmid = {35512730},
   publisher = {Nature Publishing Group},
   title = {Fault-tolerant operation of a logical qubit in a diamond quantum processor},
   volume = {606},
   url = {https://www.nature.com/articles/s41586-022-04819-6},
   year = {2022},
}

@article{Egan2021,
   author = {Laird Egan and Dripto M. Debroy and Crystal Noel and Andrew Risinger and Daiwei Zhu and Debopriyo Biswas and Michael Newman and Muyuan Li and Kenneth R. Brown and Marko Cetina and Christopher Monroe},
   doi = {10.1038/s41586-021-03928-y},
   issn = {1476-4687},
   issue = {7880},
   journal = {Nature},
   month = {10},
   pages = {281-286},
   pmid = {34608286},
   publisher = {Nature Publishing Group},
   title = {Fault-tolerant control of an error-corrected qubit},
   volume = {598},
   url = {https://www.nature.com/articles/s41586-021-03928-y},
   year = {2021},
}

@article{Acharya2023,
   author = {Rajeev Acharya and Igor Aleiner and Richard Allen and Trond I. Andersen and Markus Ansmann and Frank Arute and Kunal Arya and Abraham Asfaw and Juan Atalaya and Ryan Babbush and Dave Bacon and Joseph C. Bardin and Joao Basso and Andreas Bengtsson and Sergio Boixo and Gina Bortoli and Alexandre Bourassa and Jenna Bovaird and Leon Brill and Michael Broughton and Bob B. Buckley and David A. Buell and Tim Burger and Brian Burkett and Nicholas Bushnell and Yu Chen and Zijun Chen and Ben Chiaro and Josh Cogan and Roberto Collins and Paul Conner and William Courtney and Alexander L. Crook and Ben Curtin and Dripto M. Debroy and Alexander Del Toro Barba and Sean Demura and Andrew Dunsworth and Daniel Eppens and Catherine Erickson and Lara Faoro and Edward Farhi and Reza Fatemi and Leslie Flores Burgos and Ebrahim Forati and Austin G. Fowler and Brooks Foxen and William Giang and Craig Gidney and Dar Gilboa and Marissa Giustina and Alejandro Grajales Dau and Jonathan A. Gross and Steve Habegger and Michael C. Hamilton and Matthew P. Harrigan and Sean D. Harrington and Oscar Higgott and Jeremy Hilton and Markus Hoffmann and Sabrina Hong and Trent Huang and Ashley Huff and William J. Huggins and Lev B. Ioffe and Sergei V. Isakov and Justin Iveland and Evan Jeffrey and Zhang Jiang and Cody Jones and Pavol Juhas and Dvir Kafri and Kostyantyn Kechedzhi and Julian Kelly and Tanuj Khattar and Mostafa Khezri and Mária Kieferová and Seon Kim and Alexei Kitaev and Paul V. Klimov and Andrey R. Klots and Alexander N. Korotkov and Fedor Kostritsa and John Mark Kreikebaum and David Landhuis and Pavel Laptev and Kim Ming Lau and Lily Laws and Joonho Lee and Kenny Lee and Brian J. Lester and Alexander Lill and Wayne Liu and Aditya Locharla and Erik Lucero and Fionn D. Malone and Jeffrey Marshall and Orion Martin and Jarrod R. McClean and Trevor McCourt and Matt McEwen and Anthony Megrant and Bernardo Meurer Costa and Xiao Mi and Kevin C. Miao and Masoud Mohseni and Shirin Montazeri and Alexis Morvan and Emily Mount and Wojciech Mruczkiewicz and Ofer Naaman and Matthew Neeley and Charles Neill and Ani Nersisyan and Hartmut Neven and Michael Newman and Jiun How Ng and Anthony Nguyen and Murray Nguyen and Murphy Yuezhen Niu and Thomas E. O’Brien and Alex Opremcak and John Platt and Andre Petukhov and Rebecca Potter and Leonid P. Pryadko and Chris Quintana and Pedram Roushan and Nicholas C. Rubin and Negar Saei and Daniel Sank and Kannan Sankaragomathi and Kevin J. Satzinger and Henry F. Schurkus and Christopher Schuster and Michael J. Shearn and Aaron Shorter and Vladimir Shvarts and Jindra Skruzny and Vadim Smelyanskiy and W. Clarke Smith and George Sterling and Doug Strain and Marco Szalay and Alfredo Torres and Guifre Vidal and Benjamin Villalonga and Catherine Vollgraff Heidweiller and Theodore White and Cheng Xing and Z. Jamie Yao and Ping Yeh and Juhwan Yoo and Grayson Young and Adam Zalcman and Yaxing Zhang and Ningfeng Zhu},
   doi = {10.1038/s41586-022-05434-1},
   issn = {1476-4687},
   issue = {7949},
   journal = {Nature},
   month = {2},
   pages = {676-681},
   pmid = {36813892},
   publisher = {Nature Publishing Group},
   title = {Suppressing quantum errors by scaling a surface code logical qubit},
   volume = {614},
   url = {https://www.nature.com/articles/s41586-022-05434-1},
   year = {2023},
}

@article{Krinner2022,
   author = {Sebastian Krinner and Nathan Lacroix and Ants Remm and Agustin Di Paolo and Elie Genois and Catherine Leroux and Christoph Hellings and Stefania Lazar and Francois Swiadek and Johannes Herrmann and Graham J. Norris and Christian Kraglund Andersen and Markus Müller and Alexandre Blais and Christopher Eichler and Andreas Wallraff},
   doi = {10.1038/s41586-022-04566-8},
   issn = {1476-4687},
   issue = {7911},
   journal = {Nature},
   month = {5},
   pages = {669-674},
   pmid = {35614249},
   publisher = {Nature Publishing Group},
   title = {Realizing repeated quantum error correction in a distance-three surface code},
   volume = {605},
   url = {https://www.nature.com/articles/s41586-022-04566-8},
   year = {2022},
}

@article{Postler2022,
   author = {Lukas Postler and Sascha Heuβen and Ivan Pogorelov and Manuel Rispler and Thomas Feldker and Michael Meth and Christian D. Marciniak and Roman Stricker and Martin Ringbauer and Rainer Blatt and Philipp Schindler and Markus Müller and Thomas Monz},
   doi = {10.1038/s41586-022-04721-1},
   issn = {1476-4687},
   issue = {7911},
   journal = {Nature},
   month = {5},
   pages = {675-680},
   pmid = {35614250},
   publisher = {Nature Publishing Group},
   title = {Demonstration of fault-tolerant universal quantum gate operations},
   volume = {605},
   url = {https://www.nature.com/articles/s41586-022-04721-1},
   year = {2022},
}

@article{Ryan2022,
   author = {C. Ryan-Anderson and N. C. Brown and M. S. Allman and B. Arkin and G. Asa-Attuah and C. Baldwin and J. Berg and J. G. Bohnet and S. Braxton and N. Burdick and J. P. Campora and A. Chernoguzov and J. Esposito and B. Evans and D. Francois and J. P. Gaebler and T. M. Gatterman and J. Gerber and K. Gilmore and D. Gresh and A. Hall and A. Hankin and J. Hostetter and D. Lucchetti and K. Mayer and J. Myers and B. Neyenhuis and J. Santiago and J. Sedlacek and T. Skripka and A. Slattery and R. P. Stutz and J. Tait and R. Tobey and G. Vittorini and J. Walker and D. Hayes},
   month = {8},
   title = {Implementing Fault-tolerant Entangling Gates on the Five-qubit Code and the Color Code},
   journal = {arXiv:2208.01863v1},
   url = {https://arxiv.org/abs/2208.01863v1},
   year = {2022},
}

@article{Bedalov2024,
   author = {Matt. J. Bedalov and Matt Blakely and Peter. D. Buttler and Caitlin Carnahan and Frederic T. Chong and Woo Chang Chung and Dan C. Cole and Palash Goiporia and Pranav Gokhale and Bettina Heim and Garrett T. Hickman and Eric B. Jones and Ryan A. Jones and Pradnya Khalate and Jin-Sung Kim and Kevin W. Kuper and Martin T. Lichtman and Stephanie Lee and David Mason and Nathan A. Neff-Mallon and Thomas W. Noel and Victory Omole and Alexander G. Radnaev and Rich Rines and Mark Saffman and Efrat Shabtai and Mariesa H. Teo and Bharath Thotakura and Teague Tomesh and Angela K. Tucker},
   month = {12},
   title = {Fault-Tolerant Operation and Materials Science with Neutral Atom Logical Qubits}, 
   journal = {arXiv:2412.07670v1},
   url = {https://arxiv.org/abs/2412.07670v1},
   year = {2024},
}

@article{Singh2024,
   author = {Divyanshu Singh and Shiroman Prakash},
   month = {12},
   title = {Fault-Tolerant Implementation of the Deutsch-Jozsa Algorithm},
   journal = {arXiv: 2412.04791v1},
   url = {https://arxiv.org/abs/2412.04791v1},
   year = {2024},
}

@article{Takeda2022,
   author = {Kenta Takeda and Akito Noiri and Takashi Nakajima and Takashi Kobayashi and Seigo Tarucha},
   doi = {10.1038/s41586-022-04986-6},
   issn = {1476-4687},
   issue = {7924},
   journal = {Nature},
   month = {8},
   pages = {682-686},
   pmid = {36002485},
   publisher = {Nature Publishing Group},
   title = {Quantum error correction with silicon spin qubits},
   volume = {608},
   url = {https://www.nature.com/articles/s41586-022-04986-6},
   year = {2022},
}

@article{Takagi2022,
   author = {Ryuji Takagi and Suguru Endo and Shintaro Minagawa and Mile Gu},
   doi = {10.1038/s41534-022-00618-z},
   issn = {2056-6387},
   issue = {1},
   journal = {npj Quantum Inf},
   month = {9},
   pages = {1-11},
   publisher = {Nature Publishing Group},
   title = {Fundamental limits of quantum error mitigation},
   volume = {8},
   url = {https://www.nature.com/articles/s41534-022-00618-z},
   year = {2022},
}

@article{Kim2023,
   author = {Youngseok Kim and Andrew Eddins and Sajant Anand and Ken Xuan Wei and Ewout van den Berg and Sami Rosenblatt and Hasan Nayfeh and Yantao Wu and Michael Zaletel and Kristan Temme and Abhinav Kandala},
   doi = {10.1038/s41586-023-06096-3},
   issn = {1476-4687},
   issue = {7965},
   journal = {Nature},
   month = {6},
   pages = {500-505},
   pmid = {37316724},
   publisher = {Nature Publishing Group},
   title = {Evidence for the utility of quantum computing before fault tolerance},
   volume = {618},
   url = {https://www.nature.com/articles/s41586-023-06096-3},
   year = {2023},
}

@article{Saxena2024,
   author = {Gaurav Saxena and Thi Ha Kyaw},
   month = {9},
   title = {Error Mitigation by Restricted Evolution},
   journal = {arXiv:2409.06636v1},
   url = {https://arxiv.org/abs/2409.06636v1},
   year = {2024},
}

@article{Temme2017,
   author = {Kristan Temme and Sergey Bravyi and Jay M. Gambetta},
   doi = {https://doi.org/10.1103/PhysRevLett.119.180509},
   issn = {10797114},
   issue = {18},
   journal = {Phys. Rev. Lett. },
   month = {11},
   pages = {180509},
   pmid = {29219599},
   publisher = {American Physical Society},
   title = {Error Mitigation for Short-Depth Quantum Circuits},
   volume = {119},
   url = {https://doi.org/10.1103/PhysRevLett.119.180509},
   year = {2017},
}

@article{Huggins2021,
   author = {William J. Huggins and Sam McArdle and Thomas E. O'Brien and Joonho Lee and Nicholas C. Rubin and Sergio Boixo and K. Birgitta Whaley and Ryan Babbush and Jarrod R. McClean},
   doi = {https://doi.org/10.1103/PhysRevX.11.041036},
   issn = {21603308},
   issue = {4},
   journal = {Phys. Rev. X},
   month = {12},
   pages = {041036},
   publisher = {American Physical Society},
   title = {Virtual Distillation for Quantum Error Mitigation},
   volume = {11},
   url = {https://doi.org/10.1103/PhysRevX.11.041036},
   year = {2021},
}

@article{Cai2021,
   author = {Zhenyu Cai},
   doi = {10.22331/q-2021-09-21-548},
   issn = {2521327X},
   issue = {5},
   journal = {Quantum},
   month = {9},
   pages = {548},
   publisher = {Verein zur Förderung des Open Access Publizierens in den Quantenwissenschaften},
   title = {Quantum Error Mitigation using Symmetry Expansion},
   volume = {5},
   url = {https://quantum-journal.org/papers/q-2021-09-21-548/},
   year = {2021},
}

@article{Yoshioka2022,
   author = {Nobuyuki Yoshioka and Hideaki Hakoshima and Yuichiro Matsuzaki and Yuuki Tokunaga and Yasunari Suzuki and Suguru Endo},
   doi = {https://doi.org/10.1103/PhysRevLett.129.020502},
   issn = {10797114},
   issue = {2},
   journal = {Phys. Rev. Lett. },
   month = {7},
   pages = {020502},
   pmid = {35867434},
   publisher = {American Physical Society},
   title = {Generalized Quantum Subspace Expansion},
   volume = {129},
   url = {https://doi.org/10.1103/PhysRevLett.129.020502},
   year = {2022},
}

@article{Kim2020,
   author = {Changjun Kim and Kyungdeock Daniel Park and June Koo Rhee},
   doi = {10.1109/ACCESS.2020.3031607},
   issn = {21693536},
   journal = {IEEE Access},
   pages = {18853-188860},
   publisher = {Institute of Electrical and Electronics Engineers Inc.},
   title = {Quantum error mitigation with artificial neural network},
   volume = {8},
   year = {2020},
}

@article{Li2017,
   author = {Ying Li and Simon C. Benjamin},
   doi = {https://doi.org/10.1103/PhysRevX.7.021050},
   issue = {2},
   journal = {Phys. Rev. X},
   month = {6},
   pages = {021050},
   publisher = {American Physical Society (APS)},
   title = {Efficient Variational Quantum Simulator Incorporating Active Error Minimization},
   volume = {7},
   url = {https://doi.org/10.1103/PhysRevX.7.021050},
   year = {2017},
}

@article{cai2023,
  title = {Quantum error mitigation},
  author = {Cai, Zhenyu and Babbush, Ryan and Benjamin, Simon C. and Endo, Suguru and Huggins, William J. and Li, Ying and McClean, Jarrod R. and O'Brien, Thomas E.},
  journal = {Rev. Mod. Phys.},
  volume = {95},
  issue = {4},
  pages = {045005},
  numpages = {37},
  year = {2023},
  month = {12},
  publisher = {American Physical Society},
  doi = {10.1103/RevModPhys.95.045005},
  url = {https://link.aps.org/doi/10.1103/RevModPhys.95.045005}
}

@misc{Petz2008,
  author       = {Dénes Petz},
  title        = {},
  howpublished = {"Quantum information theory and quantum statistics". \href{https://doi.org/10.1007/978-3-540-74636-2}{Springer Berlin, Heidelberg (2007)}},
}

@article{ewout2023,
   author = {Ewout van den Berg and Zlatko K. Minev and Abhinav Kandala and Kristan Temme},
   doi = {10.1038/s41567-023-02042-2},
   issn = {1745-2481},
   issue = {8},
   journal = {Nat. Phy.},
   month = {5},
   pages = {1116-1121},
   publisher = {Nature Publishing Group},
   title = {Probabilistic error cancellation with sparse Pauli–Lindblad models on noisy quantum processors},
   volume = {19},
   url = {https://www.nature.com/articles/s41567-023-02042-2},
   year = {2023},
}

@article{Nielsen2021,
   author = {Erik Nielsen and John King Gamble and Kenneth Rudinger and Travis Scholten and Kevin Young and Robin Blume-Kohout},
   doi = {10.22331/q-2021-10-05-557},
   issn = {2521327X},
   journal = {Quantum},
   month = {10},
   pages = {557},
   publisher = {Verein zur Förderung des Open Access Publizierens in den Quantenwissenschaften},
   title = {Gate Set Tomography},
   volume = {5},
   url = {https://quantum-journal.org/papers/q-2021-10-05-557/},
   year = {2021},
}

@article{Berg2024,
   author = {Ewout van den Berg and Pawel Wocjan},
   doi = {10.22331/q-2024-12-10-1556},
   issn = {2521-327X},
   journal = {Quantum},
   month = {12},
   pages = {1556},
   publisher = {Verein zur Förderung des Open Access Publizierens in den Quantenwissenschaften},
   title = {Techniques for learning sparse Pauli-Lindblad noise models},
   volume = {8},
   url = {https://quantum-journal.org/papers/q-2024-12-10-1556/},
   year = {2024},
}

@article{Hubig2017,
   author = {C. Hubig and I. P. McCulloch and U. Schollwöck},
   doi = {https://doi.org/10.1103/PhysRevB.95.035129},
   issn = {24699969},
   issue = {3},
   journal = {Phys. Rev. B},
   month = {1},
   publisher = {American Physical Society},
   title = {Generic construction of efficient matrix product operators},
   volume = {95},
   year = {2017},
   url = {https://doi.org/10.1103/PhysRevB.95.035129},
   pages = {035129},
}

@article{Filippov2023,
   author = {Sergei Filippov and Matea Leahy and Matteo A. C. Rossi and Guillermo García-Pérez},
   month = {7},
   title = {Scalable tensor-network error mitigation for near-term quantum computing},
   journal = {arXiv:2307.11740v2},
   url = {https://arxiv.org/abs/2307.11740v2},
   year = {2023},
}

@article{Gray2018,
   author = {Johnnie Gray},
   doi = {10.21105/JOSS.00819},
   issn = {2475-9066},
   issue = {29},
   journal = {JOSS},
   month = {9},
   pages = {819},
   publisher = {The Open Journal},
   title = {quimb: A python package for quantum information and many-body calculations},
   volume = {3},
   url = {https://joss.theoj.org/papers/10.21105/joss.00819},
   year = {2018},
}

@article{Tsubouchi2023,
   author = {Kento Tsubouchi and Takahiro Sagawa and Nobuyuki Yoshioka},
   doi = {https://doi.org/10.1103/PhysRevLett.131.210601},
   issn = {10797114},
   issue = {21},
   journal = {Phys. Rev. Lett.},
   month = {11},
   pages = {210601},
   pmid = {38072608},
   publisher = {American Physical Society},
   title = {Universal Cost Bound of Quantum Error Mitigation Based on Quantum Estimation Theory},
   volume = {131},
   url = {https://doi.org/10.1103/PhysRevLett.131.210601},
   year = {2023},
}

@article{Takagi2023,
   author = {Ryuji Takagi and Hiroyasu Tajima and Mile Gu},
   doi = {https://doi.org/10.1103/PhysRevLett.131.210602},
   issn = {10797114},
   issue = {21},
   journal = {Phys. Rev. Lett.},
   month = {11},
   pages = {210602},
   pmid = {38072595},
   publisher = {American Physical Society},
   title = {Universal Sampling Lower Bounds for Quantum Error Mitigation},
   volume = {131},
   url = {https://doi.org/10.1103/PhysRevLett.131.210602},
   year = {2023},
}

@article{Fischer2024,
   author = {Laurin E. Fischer and Timothée Dao and Ivano Tavernelli and Francesco Tacchino},
   doi = {https://doi.org/10.1103/PhysRevA.109.062415},
   issn = {24699934},
   issue = {6},
   journal = {Phys. Rev. A},
   month = {6},
   publisher = {American Physical Society},
   title = {Dual-frame optimization for informationally complete quantum measurements},
   volume = {109},
   year = {2024},
   pages = {062415},
   url = {https://doi.org/10.1103/PhysRevA.109.062415},
}

@article{Malmi2024,
   author = {Joonas Malmi and Keijo Korhonen and Daniel Cavalcanti and Guillermo García-Pérez},
   doi = {https://doi.org/10.1103/PhysRevA.109.062412},
   issn = {24699934},
   issue = {6},
   journal = {Phys. Rev. A},
   month = {6},
   publisher = {American Physical Society},
   title = {Enhanced observable estimation through classical optimization of informationally overcomplete measurement data: Beyond classical shadows},
   volume = {109},
   pages = {062412},
   year = {2024},
   url = {https://doi.org/10.1103/PhysRevA.109.062412},
}

@article{Innocenti2023,
   author = {L. Innocenti and S. Lorenzo and I. Palmisano and F. Albarelli and A. Ferraro and M. Paternostro and G. M. Palma},
   doi = {https://doi.org/10.1103/PRXQuantum.4.040328},
   issn = {26913399},
   issue = {4},
   journal = {PRX Quantum},
   publisher = {American Physical Society},
   title = {Shadow Tomography on General Measurement Frames},
   volume = {4},
   year = {2023},
   pages = {040328},
   url = {https://doi.org/10.1103/PRXQuantum.4.040328},
}

@article{Watanabe2010,
  title = {Optimal Measurement on Noisy Quantum Systems},
  author = {Watanabe, Yu and Sagawa, Takahiro and Ueda, Masahito},
  journal = {Phys. Rev. Lett.},
  volume = {104},
  issue = {2},
  pages = {020401},
  numpages = {4},
  year = {2010},
  month = {1},
  publisher = {American Physical Society},
  doi = {10.1103/PhysRevLett.104.020401},
  url = {https://link.aps.org/doi/10.1103/PhysRevLett.104.020401}
}

@article{Braunstein1994,
  title = {Statistical distance and the geometry of quantum states},
  author = {Braunstein, Samuel L. and Caves, Carlton M.},
  journal = {Phys. Rev. Lett.},
  volume = {72},
  issue = {22},
  pages = {3439--3443},
  numpages = {0},
  year = {1994},
  month = {5},
  publisher = {American Physical Society},
  doi = {10.1103/PhysRevLett.72.3439},
  url = {https://link.aps.org/doi/10.1103/PhysRevLett.72.3439}
}

@article{McCourt2023,
  title = {Learning noise via dynamical decoupling of entangled qubits},
  author = {McCourt, Trevor and Neill, Charles and Lee, Kenny and Quintana, Chris and Chen, Yu and Kelly, Julian and Marshall, Jeffrey and Smelyanskiy, V. N. and Dykman, M. I. and Korotkov, Alexander and Chuang, Isaac L. and Petukhov, A. G.},
  journal = {Phys. Rev. A},
  volume = {107},
  issue = {5},
  pages = {052610},
  numpages = {5},
  year = {2023},
  month = {5},
  publisher = {American Physical Society},
  doi = {10.1103/PhysRevA.107.052610},
  url = {https://link.aps.org/doi/10.1103/PhysRevA.107.052610}
}

@article{vandenBerg2024,
  doi = {10.22331/q-2024-12-10-1556},
  url = {https://doi.org/10.22331/q-2024-12-10-1556},
  title = {Techniques for learning sparse {P}auli-{L}indblad noise models},
  author = {van den Berg, Ewout and Wocjan, Pawel},
  journal = {{Quantum}},
  issn = {2521-327X},
  publisher = {{Verein zur F{\"{o}}rderung des Open Access Publizierens in den Quantenwissenschaften}},
  volume = {8},
  pages = {1556},
  month = dec,
  year = {2024}
}

@article{jaloveckas2023,
      title={Efficient learning of Sparse Pauli Lindblad models for fully connected qubit topology}, 
      author={Jose Este Jaloveckas and Minh Tham Pham Nguyen and Lilly Palackal and Jeanette Miriam Lorenz and Hans Ehm},
      year={2023},
      journal ={arXiv:2311.11639},
      primaryClass={quant-ph},
      url={https://arxiv.org/abs/2311.11639}, 
}

@article{Orus2014,
   title={A practical introduction to tensor networks: Matrix product states and projected entangled pair states},
   volume={349},
   ISSN={0003-4916},
   url={http://dx.doi.org/10.1016/j.aop.2014.06.013},
   DOI={10.1016/j.aop.2014.06.013},
   journal={Ann. Phys.},
   publisher={Elsevier BV},
   author={Orús, Román},
   year={2014},
   month=oct, pages={117–158} }

@article{Halko2011,
author = {Halko, N. and Martinsson, P. G. and Tropp, J. A.},
title = {Finding Structure with Randomness: Probabilistic Algorithms for Constructing Approximate Matrix Decompositions},
journal = {SIAM Rev.},
volume = {53},
number = {2},
pages = {217-288},
year = {2011},
doi = {10.1137/090771806},

URL = { 
    
        https://doi.org/10.1137/090771806
    
    

},
    
    

}

@article{scott2004,
    author = {Renes, Joseph M. and Blume-Kohout, Robin and Scott, A. J. and Caves, Carlton M.},
    title = {Symmetric informationally complete quantum measurements},
    journal = {J. Math. Phys.},
    volume = {45},
    number = {6},
    pages = {2171-2180},
    year = {2004},
    month = {06},
    issn = {0022-2488},
    doi = {10.1063/1.1737053},
    url = {https://doi.org/10.1063/1.1737053},
}

@article{Efthymiou2021,
   title={Qibo: a framework for quantum simulation with hardware acceleration},
   volume={7},
   ISSN={2058-9565},
   url={http://dx.doi.org/10.1088/2058-9565/ac39f5},
   DOI={10.1088/2058-9565/ac39f5},
   number={1},
   journal={QST},
   publisher={IOP Publishing},
   author={Efthymiou, Stavros and Ramos-Calderer, Sergi and Bravo-Prieto, Carlos and Pérez-Salinas, Adrián and García-Martín, Diego and Garcia-Saez, Artur and Latorre, José Ignacio and Carrazza, Stefano},
   year={2021},
   month=dec, pages={015018} }

@article{Roffe2019,
   author = {Joschka Roffe},
   doi = {10.1080/00107514.2019.1667078},
   issue = {3},
   journal = {Contemp. Phys.},
   month = {7},
   pages = {226-245},
   publisher = {Taylor and Francis Ltd.},
   title = {Quantum Error Correction: An Introductory Guide},
   volume = {60},
   url = {http://arxiv.org/abs/1907.11157 http://dx.doi.org/10.1080/00107514.2019.1667078},
   year = {2019},
}

@article{Preskill2018,
  doi = {10.22331/q-2018-08-06-79},
  url = {https://doi.org/10.22331/q-2018-08-06-79},
  title = {Quantum {C}omputing in the {NISQ} era and beyond},
  author = {Preskill, John},
  journal = {{Quantum}},
  issn = {2521-327X},
  publisher = {{Verein zur F{\"{o}}rderung des Open Access Publizierens in den Quantenwissenschaften}},
  volume = {2},
  pages = {79},
  month = aug,
  year = {2018}
}

@misc{preskillbook,
  author       = {John Preskill},
  title        = {},
  howpublished = {"Fault-tolerant quantum computation", In \textit{Introduction to Quantum Computation and Information}. \href{https://www.worldscientific.com/doi/abs/10.1142/9789812385253_0008}{World Scientific (1998)}, pp. 213-269},
}

@article{Kandala2019,
   author = {Abhinav Kandala and Kristan Temme and Antonio D. Córcoles and Antonio Mezzacapo and Jerry M. Chow and Jay M. Gambetta},
   doi = {10.1038/s41586-019-1040-7},
   issn = {1476-4687},
   issue = {7749},
   journal = {Nature},
   month = {3},
   pages = {491-495},
   pmid = {30918370},
   publisher = {Nature Publishing Group},
   title = {Error mitigation extends the computational reach of a noisy quantum processor},
   volume = {567},
   url = {https://www.nature.com/articles/s41586-019-1040-7},
   year = {2019},
}

@article{Malone2022,
   author = {Fionn D. Malone and Robert M. Parrish and Alicia R. Welden and Thomas Fox and Matthias Degroote and Elica Kyoseva and Nikolaj Moll and Raffaele Santagati and Michael Streif},
   doi = {10.1039/D1SC05691C},
   issn = {2041-6539},
   issue = {11},
   journal = {Chemical Science},
   month = {3},
   pages = {3094-3108},
   publisher = {The Royal Society of Chemistry},
   title = {Towards the simulation of large scale protein–ligand interactions on NISQ-era quantum computers},
   volume = {13},
   url = {https://pubs.rsc.org/en/content/articlehtml/2022/sc/d1sc05691c https://pubs.rsc.org/en/content/articlelanding/2022/sc/d1sc05691c},
   year = {2022},
}

@article{Suzuki2022,
  title = {Quantum Error Mitigation as a Universal Error Reduction Technique: Applications from the NISQ to the Fault-Tolerant Quantum Computing Eras},
  author = {Suzuki, Yasunari and Endo, Suguru and Fujii, Keisuke and Tokunaga, Yuuki},
  journal = {PRX Quantum},
  volume = {3},
  issue = {1},
  pages = {010345},
  numpages = {33},
  year = {2022},
  month = {3},
  publisher = {American Physical Society},
  doi = {10.1103/PRXQuantum.3.010345},
  url = {https://link.aps.org/doi/10.1103/PRXQuantum.3.010345}
}

@article{Abughanem2024,
      title={IBM Quantum Computers: Evolution, Performance, and Future Directions}, 
      author={M. AbuGhanem},
      year={2024},
      journal ={arXiv:2410.00916},
      primaryClass={quant-ph},
      url={https://arxiv.org/abs/2410.00916}, 
}

@book{Nielsen2000,
   author = {Michael A. Nielsen and Isaac L. Chuang},
   title = {Quantum Computation and Quantum Information},
   publisher = {\href{https://doi.org/10.1017/cbo9780511976667}{Cambridge University Press}},
   year = {\href{https://doi.org/10.1017/cbo9780511976667}{2000}},
   edition = {10th Anniversary},
   doi = {10.1017/CBO9780511976667},
}

@article{Acharya2021,
   author = {Atithi Acharya and Siddhartha Saha and Anirvan M. Sengupta},
   doi = {10.1103/PHYSREVA.104.052418/FIGURES/8/THUMBNAIL},
   issn = {24699934},
   issue = {5},
   journal = {Phys. Rev. A},
   month = {11},
   pages = {052418},
   publisher = {American Physical Society},
   title = {Shadow tomography based on informationally complete positive operator-valued measure},
   volume = {104},
   url = {https://journals.aps.org/pra/abstract/10.1103/PhysRevA.104.052418},
   year = {2021},
}

@article{Sun2018,
   author = {Mingyu Sun and Michael R. Geller},
   journal = {arXive: 1810.10523},
   month = {10},
   title = {Efficient characterization of correlated SPAM errors},
   url = {https://arxiv.org/abs/1810.10523v2},
   year = {2018},
}

@article{Yu2023,
   author = {Hongye Yu and Tzu-Chieh Wei},
   journal = {arXive: 2310.18881},
   month = {10},
   title = {Efficient separate quantification of state preparation errors and measurement errors on quantum computers and their mitigation},
   url = {https://arxiv.org/abs/2310.18881v1},
   year = {2023},
}

@article{Xiong2020,
   author = {Yifeng Xiong and Daryus Chandra and Soon Xin Ng and Lajos Hanzo},
   doi = {10.1109/access.2020.3045016},
   issn = {21693536},
   journal = {IEEE Access},
   month = {12},
   pages = {228967-228991},
   publisher = {Institute of Electrical and Electronics Engineers Inc.},
   title = {Sampling Overhead Analysis of Quantum Error Mitigation: Uncoded vs. Coded Systems},
   volume = {8},
   url = {https://arxiv.org/pdf/2012.08378},
   year = {2020},
}

@article{Xiong2022,
   author = {Yifeng Xiong and Soon Xin Ng and Lajos Hanzo},
   doi = {10.1109/tcomm.2022.3144469},
   issn = {15580857},
   issue = {3},
   journal = {IEEE TCOM},
   month = {1},
   pages = {1943-1956},
   publisher = {Institute of Electrical and Electronics Engineers Inc.},
   title = {The Accuracy vs. Sampling Overhead Trade-off in Quantum Error Mitigation Using Monte Carlo-Based Channel Inversion},
   volume = {70},
   url = {https://arxiv.org/pdf/2201.07923},
   year = {2022},
}

@article{Hsieh2024,
   author = {Cheng Yun Hsieh and Hsin Ying Tsai and Yuan Hsiang Lu and James Chien Mo Li},
   doi = {10.1109/TCAD.2023.3329042},
   issn = {19374151},
   issue = {3},
   journal = {IEEE TCAD},
   month = {3},
   pages = {826-839},
   publisher = {Institute of Electrical and Electronics Engineers Inc.},
   title = {Small Sampling Overhead Error Mitigation for Quantum Circuits},
   volume = {43},
   year = {2024},
}

@article{Geller2021,
   author = {Michael R. Geller and Mingyu Sun},
   doi = {10.1088/2058-9565/ABD5C9},
   issn = {2058-9565},
   issue = {2},
   journal = {Quantum Science and Technology},
   month = {2},
   pages = {025009},
   publisher = {IOP Publishing},
   title = {Toward efficient correction of multiqubit measurement errors: pair correlation method},
   volume = {6},
   url = {https://iopscience.iop.org/article/10.1088/2058-9565/abd5c9 https://iopscience.iop.org/article/10.1088/2058-9565/abd5c9/meta},
   year = {2021}
}

@article{Lin2021,
   author = {Junan Lin and Joel J. Wallman and Ian Hincks and Raymond Laflamme},
   doi = {10.1103/PhysRevResearch.3.033285},
   issn = {26431564},
   issue = {3},
   journal = {Physical Review Research},
   month = {9},
   pages = {033285},
   publisher = {American Physical Society},
   title = {Independent state and measurement characterization for quantum computers},
   volume = {3},
   url = {https://journals.aps.org/prresearch/abstract/10.1103/PhysRevResearch.3.033285},
   year = {2021}
}

@article{Jayakumar2024,
   author = {Abhijith Jayakumar and Stefano Chessa and Carleton Coffrin and Andrey Y. Lokhov and Marc Vuffray and Sidhant Misra},
   doi = {10.22331/q-2024-07-30-1426},
   issn = {2521327X},
   journal = {Quantum},
   month = {7},
   pages = {1426},
   publisher = {Verein zur Förderung des Open Access Publizierens in den Quantenwissenschaften},
   title = {Universal framework for simultaneous tomography of quantum states and SPAM noise},
   volume = {8},
   url = {https://quantum-journal.org/papers/q-2024-07-30-1426/},
   year = {2024}
}

@article{Petersen2016,
   author = {Ian R. Petersen},
   doi = {10.2174/1874444301608010067},
   issn = {1874-4443},
   issue = {1},
   journal = {The Open Automation and Control Systems Journal},
   month = {3},
   pages = {67-93},
   publisher = {Bentham Science Publishers Ltd.},
   title = {Quantum Linear Systems Theory},
   volume = {8},
   url = {https://arxiv.org/pdf/1603.04950},
   year = {2016}
}

@article{Aseguinolaza2024,
   author = {Unai Aseguinolaza and Nahual Sobrino and Gabriel Sobrino and Joaquim Jornet-Somoza and Juan Borge},
   doi = {10.1007/S11128-024-04384-Z},
   issn = {1573-1332},
   issue = {5},
   journal = {Quantum Information Processing 2024 23:5},
   month = {5},
   pages = {1-17},
   publisher = {Springer},
   title = {Error estimation in current noisy quantum computers},
   volume = {23},
   url = {https://link.springer.com/article/10.1007/s11128-024-04384-z},
   year = {2024}
}

@article{Noiri2022,
   author = {Akito Noiri and Kenta Takeda and Takashi Nakajima and Takashi Kobayashi and Amir Sammak and Giordano Scappucci and Seigo Tarucha},
   doi = {10.1038/S41586-021-04182-Y},
   issn = {1476-4687},
   issue = {7893},
   journal = {Nature},
   month = {1},
   pages = {338-342},
   pmid = {35046603},
   publisher = {Nature},
   title = {Fast universal quantum gate above the fault-tolerance threshold in silicon},
   volume = {601},
   url = {https://pubmed.ncbi.nlm.nih.gov/35046603/},
   year = {2022}
}

@article{2020SciPy-NMeth,
  author  = {Virtanen, Pauli and Gommers, Ralf and Oliphant, Travis E. and
            Haberland, Matt and Reddy, Tyler and Cournapeau, David and
            Burovski, Evgeni and Peterson, Pearu and Weckesser, Warren and
            Bright, Jonathan and {van der Walt}, St{\'e}fan J. and
            Brett, Matthew and Wilson, Joshua and Millman, K. Jarrod and
            Mayorov, Nikolay and Nelson, Andrew R. J. and Jones, Eric and
            Kern, Robert and Larson, Eric and Carey, C J and
            Polat, {\.I}lhan and Feng, Yu and Moore, Eric W. and
            {VanderPlas}, Jake and Laxalde, Denis and Perktold, Josef and
            Cimrman, Robert and Henriksen, Ian and Quintero, E. A. and
            Harris, Charles R. and Archibald, Anne M. and
            Ribeiro, Ant{\^o}nio H. and Pedregosa, Fabian and
            {van Mulbregt}, Paul and {SciPy 1.0 Contributors}},
  title   = {SciPy 1.0: Fundamental Algorithms for Scientific
            Computing in Python},
  journal = {Nat. Methods},
  year    = {2020},
  volume  = {17},
  pages   = {261-272},
  doi     = {10.1038/s41592-019-0686-2},
  url     = {https://doi.org/10.1038/s41592-019-0686-2}
}

@article{Zhou2025,
   author = {Zeyuan Zhou and Shaun Pexton and Aleksander Kubica and Yongshan Ding},
   journal = {arXiv:2512.09863},
   month = {12},
   title = {Error Mitigation of Fault-Tolerant Quantum Circuits with Soft Information},
   url = {https://arxiv.org/pdf/2512.09863},
   year = {2025}
}

\onecolumn\newpage
\appendix

\section{Quantum Cramér-Rao Bound}
\label{app: crb}
\subsection{Noiseless observation}
We are interested in the expectation value of observable $X$ when applied to the noiseless quantum state $\rho(\bm{\theta})$. From Eq. (\ref{eq: exp X}), we have:
\begin{equation}
    f(\bm{\theta}) \equiv \left<X\right>_{\rho(\bm{\theta})} = \bm{\theta}^T \bm{x}.
\end{equation}

Given $N$ independent copies of the (identical) quantum state $\rho(\bm{\theta})$, the variance of any unbiased estimator $\hat{f}(\bm{\theta})$ measuring that state is lower-bounded by the Quantum Cramér-Rao inequality \cite{Petz2008},
\begin{equation}
    \text{Var}[\hat{f}(\bm{\theta})] \geq \frac{1}{N} \nabla_\theta f(\bm{\theta})^T \; \bm{J}(\bm{\theta})^{-1} \; \nabla_\theta f(\bm{\theta}) = \frac{1}{N} \bm{x}^T \bm{J}(\bm{\theta})^{-1} \bm{x},
\end{equation}
where $\bm{J}(\bm{\theta})$ is the Quantum Fisher Information matrix \cite{Petz2008}, whose elements are defined as
\begin{equation}
    J_{ij}(\bm{\theta}) = \frac{1}{2}\Tr{\rho(\bm{\theta}) \{L_i,L_j\}},
\end{equation}
for $\{\sbullet, \sbullet \}$ being the anti-commutator, i.e,  $\{L_i,L_j\} = L_i L_j + L_j L_i$. $L_i$ is the symmetric logarithmic derivative (SLD) operator, defined implicitly as:
\begin{equation}
    \frac{\partial \rho(\bm{\theta})}{\partial \theta_i} = \frac{1}{2} \{\rho(\bm{\theta}), L_i\} = \frac{1}{2}\left[ \rho(\bm{\theta}) L_i + L_i \rho(\bm{\theta}) \right].
    \label{eq: svd definition}
\end{equation}
It is worth mentioning that $\bm{J}(\bm{\theta})$ is symmetric, as
\begin{equation}
\begin{split}
    J_{ij}(\bm{\theta}) &= \frac{1}{2}\Tr{\rho(\bm{\theta}) (L_iL_j + L_jL_i)} = \\
    &= \Tr{\frac{\partial \rho(\bm{\theta})}{\partial \theta_i} L_j} = \Tr{\frac{\partial \rho(\bm{\theta})}{\partial \theta_j} L_i} = J_{ji}(\bm{\theta}).
\end{split}
\end{equation}
Since the SLD can be decomposed using the Pauli basis,
\begin{equation}
    L_i = \alpha_i I_{2^n} + \bm{P}^T (\bm{\phi_i} \otimes I_{2^n}),
\end{equation}
where $\alpha_i \in \mathbb{R}$ and $\bm{\phi_i} \in \mathbb{R}^{4^n -1}$, we can solve Eq. (\ref{eq: svd definition}) as follows:
\begin{equation}
     \frac{\partial \rho(\bm{\theta})}{\partial \theta_i} = \frac{1}{2^n}P_i,
     \label{eq: term 1}
\end{equation}
\begin{equation}
    \begin{split}
        \frac{1}{2} \{\rho(\bm{\theta}), L_i\} &= \frac{\alpha_i}{2^n}I_{2^n} + \frac{1}{2^n} \bm{P}^T(\bm{\phi_i} \otimes I_{2^n}) + \frac{\alpha_i}{2^n}(\bm{\theta}^T \otimes I_{2^n})\bm{P} +\\
        &+  \frac{1}{2}\frac{1}{2^n} \{(\bm{\theta}^T \otimes I_{2^n})\bm{P}, \bm{P}^T(\bm{\phi_i} \otimes I_{2^n})\} = \\
        &= \frac{\alpha_i}{2^n}I_{2^n} + \frac{1}{2^n} \bm{P}^T(\bm{\phi_i} \otimes I_{2^n}) + \frac{\alpha_i}{2^n}(\bm{\theta}^T \otimes I_{2^n})\bm{P} + \\
        &+ \frac{1}{2^n} \bm{\theta}^T \bm{\phi}_i I_{2^n} + \frac{1}{2}\frac{1}{2^n} \sum_{k,j,m} \phi_{i,k} \theta_{j} \mu_{kjm} P_m,
        \label{eq: term 2}
    \end{split}
\end{equation}
where we have used Eq. (\ref{eq: rho definition}) together with the following properties,
\begin{equation}
    \{P_i, P_j\} = 2\delta_{ij}I_{2^n} + \sum_{m=1}^{4^n-1}\mu_{ijm}P_m,
\end{equation}
\begin{equation}
    [P_i, P_j] = i\sum_{m=1}^{4^n-1}\epsilon_{ijm}P_m,
\end{equation}
\begin{equation}
    \mu_{ijm} = \frac{1}{2}\Tr{\{P_i, P_j\} P_m},
\end{equation}
\begin{equation}
    \mu_{ijm} = -\frac{i}{2}\Tr{\{P_i, P_j\} P_m},
\end{equation}
where $[\sbullet, \sbullet ]$ is the commutator, i.e  $[L_i,L_j] = L_i L_j - L_j L_i$. Note that $\mu_{ijm}$ is invariant under permutation.

Equating Eq. (\ref{eq: term 1}) and (\ref{eq: term 2}) we get:
\begin{equation}
    \alpha_i = -\bm{\theta}^T \bm{\phi}_i,
    \label{eq: alhpa_i definition}
\end{equation}
\begin{equation}
    \delta_{ij} = \phi_{i,m} + \alpha_i \theta_m + \sum_{k,j=1}^{4^n-1} \phi_{i,k} \theta_j \mu_{kjm}.
    \label{eq: delta without G}
\end{equation}

Let us now define matrix $\bm{G}(\bm{\theta})$ as:
\begin{equation}
    G_{ij} (\bm{\theta}) \equiv \frac{1}{2} \Tr{\rho(\bm{\theta})\{P_i, P_j\}} = \delta_{ij} + \sum_{m=1}^{4^n-1} \mu_{ijm} \theta_m.
    \label{eq: G definition}
\end{equation}
The sum over index $j$ in Eq. (\ref{eq: delta without G}) can be rewritten using Eq. (\ref{eq: G definition}) as follows: 
\begin{equation}
    \delta_{ij} = \alpha_i \theta_m + \sum_{k=1}^{4^n-1} \phi_{i,k} G_{km} = \alpha_i \theta_m + G_{:m}(\bm{\theta})^T \bm{\phi}_i ,
    \label{eq: delta with G}
\end{equation}
where $G_{:m}(\bm{\theta})$ is the $m$-th column of $\bm{G}(\bm{\theta})$. Eq. \ref{eq: delta with G} implies:
\begin{equation}
    \bm{\delta}_i = \alpha_i \bm{\theta} + \bm{G}_{\bm{\theta}}^T \bm{\phi}_i.
\end{equation}
Considering Eq. (\ref{eq: alhpa_i definition}),  the final expression for $\bm{\phi}_i$ is given by:
\begin{equation}
    \bm{\phi}_i = (\bm{G}_{\bm{\theta}}^T - \bm{\theta} \bm{\theta}^T)^{-1} \bm{\delta}_i.
\end{equation}
This leads to the following identity:
\begin{equation}
    J_{ij}(\bm{\theta}) = \phi_{ij}(\bm{\theta}).
\end{equation}
Therefore,
\begin{equation}
    \bm{J}(\bm{\theta}) = \bm{\Phi}(\bm{\theta}) = (\bm{G}_{\bm{\theta}}^T - \bm{\theta} \bm{\theta}^T)^{-1}.
\end{equation}
Thus, the QCRB can be expressed as:
\begin{equation}
    \text{Var}[\hat{f}(\bm{\theta})] \geq \frac{1}{N} (\bm{x}^T \bm{G}_{\bm{\theta}}\bm{x} - f(\bm{\theta})^2).
    \label{eq: qcrb}
\end{equation}

It is interesting to estimate the intrinsic variance of the expectation value of any target observable $X$ over $N$ (identical) copies of the state $\rho(\bm{\theta})$ in the absence of noise,
\begin{equation}
\begin{split}
    \text{Var}[\left<X\right>_{\rho(\bm{\theta})}] &= \frac{1}{N} (\left< X^2 \right>_{\rho(\bm{\theta})} - \left< X \right>_{\rho(\bm{\theta}}^2) \\
    &= \frac{1}{N} ( \Tr{\rho(\bm{\theta}) X^2} - \Tr{\rho(\bm{\theta}) X}^2) \\
    &= \frac{1}{N} (\bm{x}^T \bm{G_{\bm{\theta}}} \bm{x} - f(\bm{\theta})^2).
    \label{eq: noiseless variance of X}
\end{split}
\end{equation}

In such a noiseless scenario, both Eq. (\ref{eq: qcrb}) and Eq. (\ref{eq: noiseless variance of X}) are equal. This means that the estimator $\hat{f}(\bm{\theta})$ is optimal, in the sense that it saturates the QCRB. Thus, there cannot exist any other unbiased estimator that reduces the variance of the target expectation value $\left<X\right>_{\rho(\bm{\theta})}$.

\subsection{Noisy observation}
Let us consider Eq. (\ref{eq: Y in paulis}) and Eq. (\ref{eq: trace of Y}). The adjoint noisy quantum channel is given by
\begin{equation}
        \mathcal{E}^\dagger(\rho(\bm{\theta})) = \frac{1}{2^n} \left( \mathcal{E}^\dagger(I_{2^n}) + \sum_{i=1}^{4^n-1} \theta_i \mathcal{E}^\dagger(P_i) \right) = \rho(\bm{A} ^\dagger \bm{\theta} + \bm{c}^\dagger),
\end{equation}
where $A_{ij}^\dagger = \frac{1}{2^n} \Tr{P_i \mathcal{E}^\dagger(P_j)}$ and $c_{i}^\dagger = \frac{1}{2^n} \Tr{I_{2^n} \mathcal{E}^\dagger(P_i)}$.
When this map is applied to the $Y$ observable, we get
\begin{equation}
    \mathcal{E}^\dagger(Y) = \bm{P}^T (\bm{A}^\dagger \bm{y} \otimes I_{2^n}) + (y_0 + (\bm{c}^\dagger)^T \bm{y}) \cdot I_{2^n}.
\end{equation}
Note that 
\begin{equation}
    A^\dagger_{ij} = \frac{1}{2^n}\Tr{P_i \mathcal{E}^\dagger(P_j)} = \frac{1}{2^n}\Tr{P_j \mathcal{E}(P_i)} = A_{ji}
\end{equation}
\begin{equation}
    c^\dagger_i = \frac{1}{2^n}\Tr{I_{2^n} \mathcal{E}^\dagger(P_i)} = \frac{1}{2^n}\Tr{\mathcal{E}^\dagger(P_i) I_{2^n}} = \frac{1}{2^n}\Tr{P_i \mathcal{E}(I_{2^n})} = c_i,
\end{equation}
which means that $\bm{A}^\dagger = \bm{A}^T$ and $\bm{c}^\dagger = \bm{c}^T$.

Solving $\mathcal{E}^\dagger(Y) = X$ we find
\begin{equation}
    \bm{y} =\bm{A}^{-T} \bm{x} ,
\end{equation}
\begin{equation}
    y_0 = - \bm{c}^T \bm{y} = - \bm{c}^T \bm{A}^{-T} \bm{x}.
\end{equation}

Measuring the $Y$ observable defines a new unbiased estimator $\hat{f'}(\bm{\theta}')$ for its expectation value, where $f'(\bm{\theta}')$ is defined from (\ref{eq: trace of Y}):
\begin{equation}
    f'(\bm{\theta}') \equiv \left< Y \right>_{\mathcal{E}(\rho(\bm{\theta}))} = (\bm{A}\bm{\theta} + \bm{c})^T \bm{y} + y_0 = \bm{\theta}'^T \bm{y} + y_0 ,
\end{equation}
where $\bm{\theta}' = \bm{A}\bm{\theta} + \bm{c}$. Notice that this estimator gives the same result as $f(\bm{\theta})$:
\begin{equation}
    f'(\bm{\theta}') = \left< Y \right>_{\mathcal{E}(\rho(\bm{\theta}))} = (\bm{A} \bm{\theta})^T \bm{y} + \bm{c}^T \bm{y} + y_0 = \bm{\theta}^T \bm{x} = \left< X \right>_{\rho(\bm{\theta}))} =  f(\bm{\theta}).
\end{equation}

In this case, the QCRB for the estimator $\hat{f'}(\bm{\theta}')$ of the expectation value of $Y$ arising from the noisy state $\mathcal{E}(\rho(\bm{\theta}))$ is
\begin{equation}
\begin{split}
    \text{Var}[f'(\bm{\theta}')] &\geq \frac{1}{N} \nabla_{\theta'} f'(\bm{\theta'}) \; \bm{J}(\bm{\theta}')^{-1} \; \nabla_{\bm{\theta}'} f'(\bm{\theta}') \\
    &= \frac{1}{N} \bm{y}^T  \bm{J}(\bm{\theta}')^{-1} \bm{y} \\
    &= \frac{1}{N} \bm{x}^T \bm{A}^{-1} (\bm{G}_{(\bm{A}\bm{\theta} + \bm{c})} + (\bm{A}\bm{\theta} + \bm{c}) (\bm{A}\bm{\theta} + \bm{c})T) \bm{A}^{-T} \bm{x} \\
    &= \frac{1}{N} (\bm{x}^T \bm{A}^{-1} \bm{G}_{(\bm{A}\bm{\theta} + \bm{c})} \bm{A}^{-T} \bm{x} - (f(\bm{\theta}) - y_0)^2). 
    \label{eq: noisy qcrb}
\end{split}
\end{equation}
The intrinsic variance for the expectation value of $Y$ over $N$ (identical) copies of the noisy state $\mathcal{E}(\rho(\bm{\theta}))$ is given by:
\begin{equation}
    \begin{split}
    \text{Var}[Y] &= \frac{1}{N} (\left< Y^2 \right>_{\mathcal{E}(\rho(\bm{\theta}))} - \left< Y \right>_{\mathcal{E}(\rho(\bm{\theta}))}^2) \\
    &= \frac{1}{N} (\Tr{\mathcal{E}(\rho(\bm{\theta})) Y^2} - \Tr{\mathcal{E}(\rho(\bm{\theta})) Y}^2) \\
    &= \frac{1}{N} (\bm{x}^T \bm{A}^{-1} \bm{G_{\bm{A}\bm{\theta} + \bm{c}}} \bm{A}^{-T} \bm{x} - (f(\bm{\theta}) - y_0)^2).
    \label{eq: variance of Y}
\end{split} 
\end{equation}
As for the single sample case, both Eq. (\ref{eq: noisy qcrb}) and Eq. (\ref{eq: variance of Y}) coincide. Thus, the expectation value $\left< Y \right>_{\mathcal{E}(\rho(\bm{\theta}))}$ saturates the QCRB. This proves the optimality of the considered estimator.

\section{Pauli Transfer Matrix Representation}
\label{app: PTM}
Since the conventional set of Pauli operators $\{ \sigma_0, \sigma_1, \sigma_2, \sigma_3\}$ forms a basis in the linear space of operators acting on the 2-dimensional Hilbert space for a single qubit, any operator $A$ is uniquely determined by a 4-dimensional vector (rank-1 tensor $\mathfrak{a}$) with components
\begin{equation}
    \mathfrak{a}_i = \frac{1}{\sqrt{2}}\Tr{A \sigma_i}, \quad i = 0,1,2,3.
\end{equation}
Thus, we have:
\begin{equation}
    A = \frac{1}{\sqrt{2}}\sum_{i=0}^3 \mathfrak{a}_i \sigma_i.
\end{equation}
Using this notation, the Hilbert-Schmidt scalar product of two operators $A$ and $B$ becomes
\begin{equation}
    \Tr{A^\dagger B} = \mathfrak{a}^\dagger \mathfrak{b},
\end{equation}
that is, the conventional scalar product of vectors $\mathfrak{a}$ and $\mathfrak{b}$.

A linear map $\mathcal{E}$ on the space of qubit operators is uniquely defined by the $4 \times 4$ matrix (rank-2 tensor) $\mathfrak{E}$ with elements
\begin{equation}
    \mathfrak{E}_{ij} = \frac{1}{2}\Tr{\sigma_i \mathcal{E}(\sigma_j)}, \quad i,j = 0,1,2,3.
\end{equation}

This representation is very efficient in the context of the composition of maps. For example, the operator $\mathcal{E}(A)$ corresponds to the product $\mathfrak{E} \mathfrak{a}$, while the composition of two maps $\mathcal{C} = \mathcal{A} \circ \mathcal{B}$ becomes the matrix product of each of the individual PTM parts, $\mathfrak{C} = \mathfrak{A} \cdot \mathfrak{B}$. The same holds for the tensor product of two maps $\mathcal{C} = \mathcal{A} \otimes \mathcal{B}$, which simplifies to $\mathfrak{C} = \mathfrak{A} \otimes \mathfrak{B}$

\section{Extra expectation values of other observables}

\begin{figure}
  \centering
  \begin{subfigure}[b]{0.45\textwidth}
    \centering
    \includegraphics[width=\textwidth]{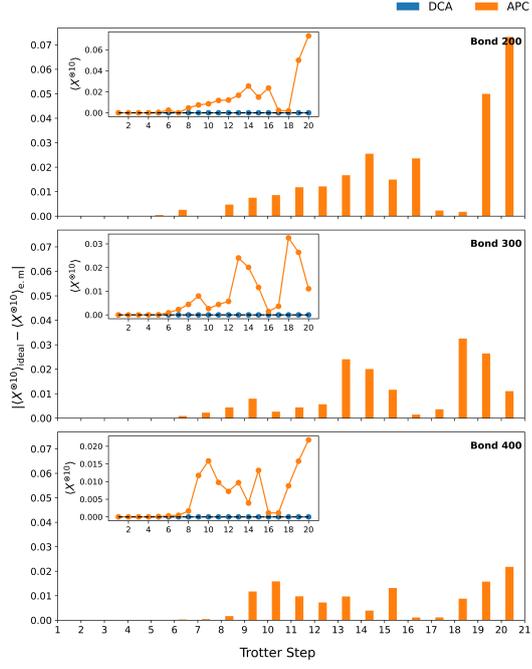}
    \caption{}
    \label{fig:image1}
  \end{subfigure}
  \hfill
  \begin{subfigure}[b]{0.45\textwidth}
    \centering
    \includegraphics[width=\textwidth]{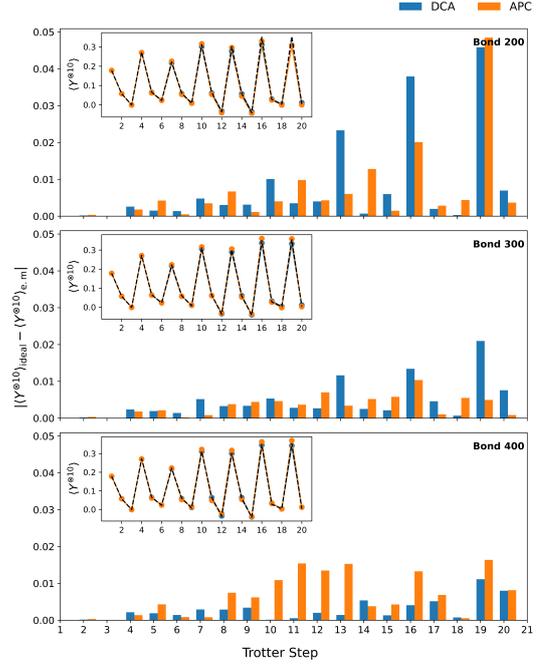}
    \caption{}
    \label{fig:image2}
  \end{subfigure}
  \vskip\baselineskip
  \begin{subfigure}[b]{0.45\textwidth}
    \centering
    \includegraphics[width=\textwidth]{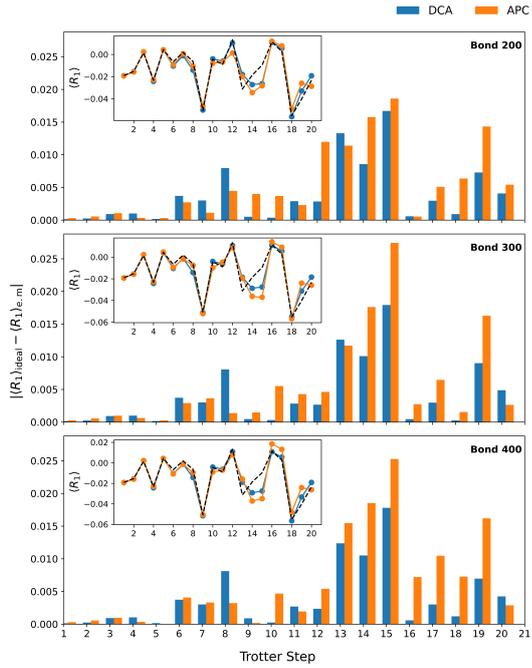}
    \caption{}
    \label{fig:image3}
  \end{subfigure}
  \hfill
  \begin{subfigure}[b]{0.45\textwidth}
    \centering
    \includegraphics[width=\textwidth]{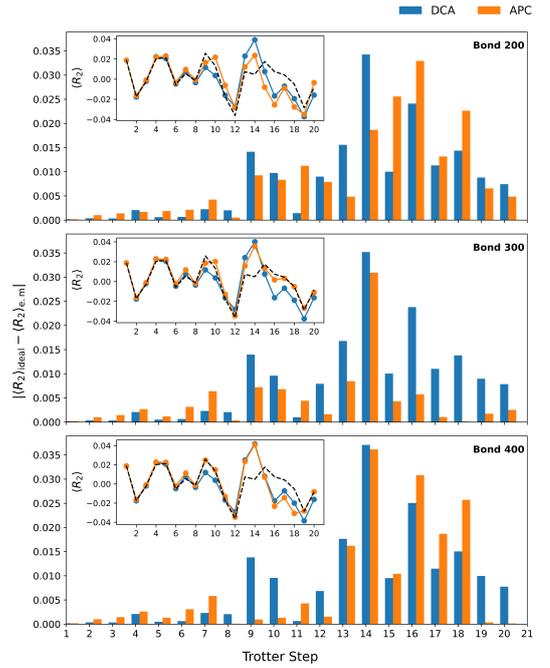}
    \caption{}
    \label{fig:image4}
  \end{subfigure}

  \caption{Absolute difference between the ideal and noise mitigated outcome \( |\left<O \right>_\text{ideal} - \left<O \right>_\text{e.m.}| \) as a function of the trotter step, considering the Dominant Component Approximation (blue) and all Pauli components (orange) analytically. We compute it for different bond dimension truncations: 200, 300, 400. The inset indicates the mitigated outcomes compared to the ideal one (dashed lines) as a function of the Trotter step. \( O = \) (a) \( X^{\otimes 10} \), (b) \( Y^{\otimes 10} \), (c) \( R_1 = Y \otimes Z \otimes Z \otimes I \otimes Y \otimes X \otimes Z \otimes I \otimes Z \otimes Z \), (d) \( R_2 = I \otimes I \otimes Z \otimes I \otimes X \otimes I \otimes X \otimes X \otimes I \otimes X \).}
  \label{fig:4images}
\end{figure}

In this section, we numerically compare the performance of the DCA with that of the APC for a set of 4 Pauli observables. We consider two high-weight Pauli observables $X^{\otimes n}$ and $Y^{\otimes n}$, as they are more difficult to mitigate using the standard Classical Shadows \cite{Acharya2021}. We also consider two random Pauli strings, $R_1 = Y \otimes Z \otimes Z \otimes I \otimes Y \otimes X \otimes Z \otimes I \otimes Z \otimes Z$ and $R_2 = I \otimes I \otimes Z \otimes I \otimes X \otimes I \otimes X \otimes X \otimes I \otimes X$. Fig. \ref{fig:4images} shows that DCA performs better than APC in most Trotter steps and tends to achieve better results at longer depths. This indicates that considering all Pauli terms of the surrogate $\hat{Y}$ tends to increase error during the mitigation procedure, as the bond dimension used during MPO and MPS truncation is inefficient to capture all relevant information at such depths.   

\end{document}